\documentclass[11pt]{article}

\usepackage[utf8]{inputenc}

\usepackage{subfig}
\usepackage{float}
\usepackage{fancyhdr}
\usepackage{amssymb}
\usepackage{algorithm}
\usepackage{algpseudocode} 
\usepackage{amsfonts}
\usepackage{graphicx}
\usepackage{hyperref}
\usepackage{mathtools}

\newtheorem{thm}{Theorem}
\newtheorem{theorem}{Theorem}

\newtheorem{lemma}[thm]{Lemma}
\newtheorem{definition}[thm]{Definition}
\newtheorem{proposition}[thm]{Proposition}

\newtheorem{claim}[thm]{Claim}

\newtheorem{assumption}[thm]{Assumption}
\newenvironment{proof}{\noindent\bf{Proof }\rm}{\hfill$\blacksquare$\bigskip}

\newcommand{\GSS}{\mathsf{GetOptimalBatches}}

\newcommand{\eps}{\epsilon}

\newcommand{\OPT}{\textsf{OPT}} 
\newcommand{\hOPT}{\widehat{\textsf{OPT}}} 
\newcommand{\cost}{\text{cost}} 
\newcommand{\opt}{\text{cost}(\textsf{OPT})} 

\newcommand{\QQ}{{\mathbb{Q}}}

\def\cL{{\cal{L}}}

\newcommand{\cA}{\mathcal{A}}

\usepackage{setspace}

\title{Online Flexible Busy Time Scheduling on Heterogeneous Machines}

\author{Adrian Calinescu\thanks{\texttt{adriancyay@gmail.com}} \and Gruia C\u alinescu\thanks{\texttt{gcali9999@gmail.com}. Research Institute of the University of Bucharest.  Work performed in part while at Northwestern University and at Illinois Institute of Technology.} \and Sami Davies\thanks{\texttt{samidavies@berkeley.edu}. Department of EECS and the Simons Institute for the Theory of Computing, UC Berkeley. Supported by an NSF Computing Innovation Fellowship while at Northwestern University.} \and Samir Khuller\thanks{\texttt{samir.khuller@northwestern.edu}. Department of Computer Science, Northwestern University. Work supported by (IDEAL) NSF 2216970.} \and Shirley Zhang \thanks{\texttt{szhang2@g.harvard.edu}. Paulson School of Engineering and Applied Sciences, Harvard University. Supported by an NSF Graduate Research Fellowship.}}

\begin{document}
\maketitle

\begin{abstract}
We study the online busy time scheduling model on heterogeneous machines.
In our setting, jobs with uniform processing time arrive online with a deadline that becomes known to the algorithm at the job's arrival time.
An algorithm has access to machines, each with different associated capacities and costs. The goal is to schedule jobs on machines by their deadline, so that the total cost incurred by the scheduling algorithm is minimized. 
While busy time scheduling has been well-studied, relatively little is known when machines are heterogeneous (i.e., have different costs and capacities), despite this natural theoretical generalization being the most practical model for clients using cloud computing services. 

We make significant progress in understanding this model by designing a deterministic online algorithm
with competitive ratio $8(2p-1)/p < 16$
when all jobs have uniform processing time $p$. 
A randomized version of this algorithm is $4(2p-1)/(p \ln 2)$-competitive against an oblivious adversary.
For unit-processing-time jobs, we give lower bounds of $4$ and $e$  on the competitive ratio
of deterministic and randomized online algorithms, respectively.
For unit-processing-time jobs with agreeable deadlines,
we provide a deterministic $2$-competitive online algorithm
and a matching lower bound.
\end{abstract}

\newpage

\section{Introduction}
Busy time scheduling is an energy minimization model that is well-studied due to its applicability to cloud computing and its relationship to energy usage \cite{KSST15, chang2014lp}.
In this model,
$n$ jobs arrive online, each with a deadline. 
A machine of \emph{type} $k$, for $k \in \mathbb{Z}_{\geq 0}$,
has cost $c_k \geq 1$ and capacity $B_k \in \mathbb{Z}_{\geq 1}$,
where the cost indicates how much the algorithm must pay for each timeslot 
where the machine is executing at least one job,
and the capacity dictates how many jobs the machine can execute
during each timeslot.
There are an unlimited number of machines of each type.
For each job $j$, we assume that its release time $r_j$ and deadline $d_j$
become known to the algorithm when the job is released.
The release time is the same time that the job ``arrives'' online, 
in the sense that the algorithm learns about the existence of the job.

We consider the setting where jobs have uniform processing time
$p \in \mathbb{Z}_{\geq 1}$ (so for all jobs $j$, $d_j-r_j \geq p - 1$), 
and all release times and deadlines are
integral\footnote{In most scenarios, these numbers can be scaled to be integers, and our results essentially hold.}.
Note that in our model, a unit-processing-time 
job with $r_j = d_j$ may be feasibly executed at timeslot $r_j$. Jobs can be assigned to a machine that is already processing jobs, provided that there is spare
capacity, and this may result in the machine running for a longer time.
The goal is to find a non-preemptive schedule completing all jobs by their deadlines with minimum total  cost.

Busy time scheduling is, in part, motivated by applications in cloud computing. 
Providers such as Amazon Web Services (AWS), Google Cloud, and Microsoft Azure rent virtual machines, 
each of which has some fixed cost and processing power. 
Customers are charged by how many working server hours they use, 
and therefore they can minimize their costs by making smart choices about which virtual machines they rent.
Busy time scheduling captures the difficulty that customers face in deciding whether they should greedily use cheap servers that only complete the jobs imminently due, or whether they should pay for a more expensive server to complete 
additional jobs that are waiting to be processed.

One interpretation of the problem with unit-processing-time
jobs is as a model of a shuttle scheduling problem. 
Suppose passengers arrive at an airport parking lot,
not in walking distance to the terminal, and tell a dispatcher when their flight is. The shuttle company has shuttles of varying sizes that incur different costs. The company needs to carry all passengers to the terminal, but would like to minimize its cost in doing so.

There has been a surge of theoretical interest in busy time scheduling over the past decade or so, with many variations on the model having been studied. Almost all of the work is for the case of \emph{homogeneous} machines, i.e., when machines are identical with the same costs and capacities.
We discuss the different variations of busy time scheduling on homogeneous machines in Subsection \ref{sec: rel-work}.
Much less is known about the setting when machines are heterogeneous,
despite this being (arguably) the most practical setting. Most data centers 
consist of machines of different types, built over time, and these in general have different costs and capacities.

In the special case of interval jobs\footnote{This is also called the 
\emph{inflexible} setting, and interval jobs are also called \emph{rigid}.}, there is no flexibility as to when a job is executed; each job has to instantly be scheduled upon arrival. 
This lack of flexibility can make the resulting cost  of any schedule extremely high---for example, if single jobs arrive one after the other and we schedule them all separately, we might pay a huge cost compared to a schedule that has the flexibility to wait and bundle a group of jobs together.
Ren and Tang \cite{Ren-tang} studied the problem of inflexible jobs
with \emph{heights}, where each job may take up non-unit space on each machine, on heterogeneous machines when the normalized cost-per-capacity rate is either monotonically increasing or decreasing
as the machine capacity increases.
They developed an $O(1)$-approximation algorithm (with constants 9 and 14, respectively) in the offline setting
and a $\Theta(\mu)$-competitive algorithm in the online setting, 
where $\mu$ is the ratio of the maximum job processing time to the minimum
job processing time. 
Later, Liu and Tang \cite{liu2021analysis} proved that one can obtain the same guarantees (up to a constant factor), but without the restrictions on the cost-per-capacity rate. 
They also provide matching lower bounds for both settings.

We are the first to study the online setting when jobs are \emph{flexible}, i.e., not interval, and machines are heterogeneous.
However, there is already a lower bound on the competitive ratio of
$\Omega(\sqrt{\log \mu})$ \cite{azar2019tight}\footnote{This result holds in the much simpler setting  where jobs are inflexible and machines are homogeneous.}. 
In order to study the problem without being totally restricted by this lower bound, 
we must be willing to make some kind of trade-off; 
we choose to give up the generality of arbitrary job processing times and heights.
By focusing on uniform-processing-time jobs, 
we make the problem more tractable, while still maintaining the core theoretical difficulty of the setting.
When jobs are flexible, an algorithm must make the additional decision of \emph{when} in a job's active interval it should be scheduled, and on heterogeneous machines, the algorithm must 
make the additional decision on whether to pay more now to process large batches, or pay less now and defer other jobs' completions.
At a high level,
the ``online'' and ``flexible'' aspects of the setting 
incentivize the algorithm to make decisions lazily, 
while the ``heterogeneous'' aspect punishes the algorithm for making decisions too late by forcing the algorithm to use some very expensive resource.

\subsection{Our results}
We note that an oblivious adversary must choose the input in advance of any of the algorithm's choices, while an adaptive adversary is aware of the algorithm's choices at each step and can choose the online input as a function of the algorithm's current state.

Our main result is the following two related theorems.
\begin{theorem} \label{thm: deterministic}
There is an $8(2p-1)/p$-competitive deterministic online algorithm for Uniform-processing-time Busy Time Scheduling on 
Heterogeneous Machines
with running time $O(n \log n)$, for $n$ the total number of jobs and $p$ the processing time of any job.
\end{theorem}
Observe that the competitive ratio is at most $16$ for any $p$, while for
unit-processing-time jobs ($p=1$) the competitive ratio is $8$.
\begin{theorem} \label{thm: randomized}
In the oblivious adversary model,
there is a $4(2p-1)/(p \ln 2 ) $-competitive randomized  online algorithm for 
Uniform-processing-time Busy Time Scheduling on Heterogeneous Machines
with running time $O(n \log n)$, for $n$ the total number of jobs and $p$ the processing time of any job.
\end{theorem}
Observe that the competitive ratio is at most $8/\ln 2 < 11.55$ for any $p$, while for unit-processing-time jobs 
the competitive ratio is $4/\ln 2 < 5.78$.

The main technical challenge
is handling nested intervals of jobs, that is, when two jobs $j_1$ and $j_2$ may have $r_{j_1} < r_{j_2}$ and $d_{j_1} > d_{j_2}$. 
For this, our core algorithm is given by Lemma \ref{lem: main} in Section \ref{sec: lbp+proof}  
and is designed with the sophisticated lower bound of Lemma \ref{lem: pvalid-assgn-lb} in mind.

Theorems \ref{thm: deterministic} and \ref{thm: randomized} are obtained in a rather straightforward manner from Lemma \ref{lem: main}.
The setting of unit-processing-time jobs captures the essence of the algorithm and proof strategy for general uniform-processing-time jobs,
with a cleaner presentation. To read the algorithm of Lemma \ref{lem: main} for unit-processing-time jobs, we refer to the earlier (conference) version
of this paper \cite{CDKZ24}. Our next theorems apply to unit-processing-time jobs.

\begin{theorem} \label{thm: lb_nested_deterministic}
For any $\epsilon > 0$,
no deterministic online algorithm can have competitive ratio $4 - \epsilon$ for Unit-processing-time
Busy Time Scheduling on Heterogeneous Machines.
This holds even when any two job intervals are nested or disjoint.
\end{theorem}

The proof of the above theorem uses results and  techniques from \cite{SLLA25}. 
Below $e \approx 2.718$ is Euler's number. Also, using ideas from  \cite{SLLA25}, we obtain:

\begin{theorem} \label{thm: lb_nested_randomized}
For any $\epsilon > 0$,
no randomized online algorithm can have competitive ratio $e - \epsilon$ for 
Unit-processing-time Busy Time Scheduling on Heterogeneous Machines, against an oblivious adversary.
 This holds even when any two job intervals are nested or disjoint.
\end{theorem}

Several straightforward procedures give $O(1)$-competitive algorithms when jobs have
\emph{agreeable}\footnote{Also called {\em compatible},
for example in \cite{Lawler1991ADP}, or {\em proper}, for example in \cite{CL19}.
}
deadlines,
meaning that if job $j$ arrives before job $j'$,
then the deadline of job $j'$ cannot be before that of $j$.
For unit-processing-time jobs with agreeable deadlines, we can improve our deterministic
competitive ratio from 8 to 2 by using a much simpler algorithm. 
Note that this result does not extend to the uniform-processing-time setting.

\begin{theorem} \label{thm: agreeable}
If jobs have agreeable deadlines,  there is a deterministic online
$2$-competitive algorithm for 
Unit-processing-time Busy Time Scheduling on Heterogeneous Machines
with running time $O( n K + n \log n)$, for $n$ the total number of jobs
and $K$ the number of distinct machine types.
\end{theorem}

Thus having only agreeable deadlines makes the problem of 
online Busy Time Scheduling on Heterogeneous Machines easier,
in view of Theorems \ref{thm: lb_nested_deterministic} 
and \ref{thm: lb_nested_randomized}.
The above result is complemented by a matching lower bound on the competitive ratio:

\begin{theorem}\label{thm: lb}
For any $\epsilon > 0$,
no deterministic online algorithm can have competitive ratio $2 - \epsilon$ for 
Unit-processing-time Busy Time Scheduling on Heterogeneous Machines with Agreeable Deadlines.
\end{theorem}

Theorems \ref{thm: deterministic}, \ref{thm: agreeable}, and \ref{thm: lb} were announced in the earlier (conference) version of this paper \cite{CDKZ24}.

\subsection{Related work}\label{sec: rel-work}

\textbf{Busy Time Scheduling for Arbitrary-processing-time jobs on Homogeneous Machines}\footnote{Results are for machines with finite capacities, unless explicitly stated otherwise.}.
Even scheduling interval jobs (which is easier than scheduling flexible jobs) on homogeneous machines
is \textsf{NP}-hard \cite{winkler2003wavelength}.
Khandekar et al. \cite{KSST15} showed a 5-approximation for busy time scheduling of flexible jobs with arbitrary heights. 
When jobs have uniform heights, Chang et al. \cite{chang2014lp} showed that algorithms developed by Alicherry and Bhatia \cite{alicherry2003line} and  Kumar and Rudra \cite{kumar2005approximation} are 4-approximations, and gave an improved algorithm that obtains a 3-approximation.

In the online setting when jobs are flexible, Koehler and Khuller \cite{koehler2017busy} give a 5-competitive algorithm
for the Busy Time problem when machines have infinite capacities and obtain an $O(\log \mu)$ competitive ratio 
when machines have finite capacities. Also in the infinite capacity setting, Ren and Tang \cite{ren2017online} give an algorithm with a competitive ratio of $4 + 2\sqrt{2}$, and complement this result with a lower bound on the competitive ratio of $\frac{\sqrt{5}+1}{2}$, which holds for any deterministic online scheduler. One variant of the online busy time problem considers the setting when job processing times are unknown at arrival time \cite{li2014dynamic, ren2016competitiveness}. Notably, the $\Theta(\mu)$-competitive algorithm and matching lower bound of Ren and Tang \cite{Ren-tang} hold for this setting as well. In our work, job processing times
are uniform, and therefore known at arrival. 

On homogeneous machines with interval jobs, Azar  and Vainstein~\cite{azar2019tight} developed an online algorithm with tight approximation factor 
$\Theta(\sqrt{\log \mu})$.
Improvements are known, but these involve certain assumptions about the knowledge of the profile of jobs that have not yet arrived \cite{buchbinder2021online}. 
It is worth noting that
on homogeneous machines, one can find the optimal schedule in the online or offline setting if jobs have unit processing time by using the algorithm by Finke et al. \cite{FINKE2008556}\footnote{The algorithm with a different proof of optimality
appeared previously in work by Bodlaender and Jansen \cite{BODLAENDER199593}.}.

\textbf{Power management and batch scheduling.}
Much work in scheduling theory has focused on energy minimization or power management, and busy time scheduling is one such model \cite{yao1995scheduling, irani2005algorithmic, irani2007algorithms, chan2007energy, buchbinder2008online, lin2012dynamic, antoniadis2020parallel}.
The models that are most relevant to ours study energy minimization
when the machine(s) have some cost to turn on or off \cite{chang2014lp, davies2022balancing, CW21, augustine2008optimal}. In batch scheduling problems,  machines can run groups of jobs simultaneously (also called active time scheduling). A series of papers develop approximation algorithms in the preemptive model \cite{ChangGK14,KumarK18,CaoFLMRU22}.
Classic objectives include throughput maximization \cite{bar2009throughput} and functions of flow time and makespan \cite{im2013online,WANG201737}.

\textbf{Capacitated covering problems.}
For $d >1$, $d$-dimensional Capacitated Rectangle Stabbing
is \textsf{NP}-hard, although there do exist approximation algorithms \cite{hassin1991approximation, gaur2002constant, even2008algorithms}. 
When $d=1$, algorithms for Capacitated Rectangle Stabbing solve the offline 
version of Unit-processing-time Busy Time Scheduling on Heterogeneous Machines;
rectangles are one-dimensional $x-$axis aligned intervals that correspond to jobs' arrival time, deadline intervals, and stabbers can be equipped with costs and capacities to correspond to a batch of jobs to be scheduled on a machine in a single timeslot.
This 1-dimensional version can be solved exactly via a dynamic programming algorithm \cite{even2008algorithms}, which is based on the influential work of Baptiste \cite{baptiste2006scheduling}.

Capacitated versions of covering problems related to vertex cover,
facility location, and other scheduling problems have also been studied \cite{chuzhoy2006covering, saha2012set, bansal2014geometry, carnes2015primal}.
Several online problems also have a similar flavor to ours, 
as an algorithm has to make a choice between different options, such as the ski-rental problem with multiple discount options \cite{zhang2011ski,SLLA25}, the parking permit problem with multiple durations \cite{meyerson2005parking}, and capacitated interval coloring \cite{epstein2009online}.

\subsection{Preliminaries}\label{sec:prelims}

As a warm-up, in Section \ref{sec: agreeable} we discuss our $2$-competitive deterministic 
online algorithm for unit-processing-time jobs with agreeable deadlines
(Theorem \ref{thm: agreeable}) and the matching lower bound (Theorem \ref{thm: lb}).
We also discuss some of the difficulties encountered with obtaining constant-factor algorithms with non-agreeable deadlines. 

Then in Section \ref{sec: lbp+proof}, we consider uniform-processing-time
jobs with general interval structures.  
We introduce our core lemma for proving
Theorems \ref{thm: deterministic} and \ref{thm: randomized}, Lemma \ref{lem: main},
and its preliminaries, as well as how Theorems \ref{thm: deterministic} and \ref{thm: randomized} follow from Lemma \ref{lem: main}.
We prove our sophisticated lower bound on the optimal solution's cost in Subsection \ref{sec: plower-bound}.
Then the algorithm of Lemma \ref{lem: main} appears in  Subsection \ref{sec: palgs}.

The lower bound of 4 on the competitive ratio of any deterministic 
online algorithm 
(Theorem \ref{thm: lb_nested_deterministic})
is provided to complement these algorithmic results in Section \ref{sec: adversary-4},
together with the lower bound of $e$ on 
the competitive ratio of any randomized online algorithm (Theorem \ref{thm: lb_nested_randomized}).

Appendix \ref{sec: app-algs} describes some natural algorithms that fail to obtain a constant competitive ratio,
while Appendix \ref{sec: app-tight-lb} shows that the (core) lower bound of Subsection \ref{sec: plower-bound} is tight, even for unit-processing-time jobs.

The formal set-up and notation throughout the main body of the paper is as follows.
Uniform-processing-time jobs arrive online, and  form a set  denoted by $J$,
with $|J|=n$. As above, $p \in \mathbb{Z}_{\geq 1}$ denotes the 
processing time of each job.
Job $j \in J$ is equipped with integral release time $r_j$ and integral deadline $d_j$.
A machine of type $k \in \mathbb{Z}_{\geq 0}$ can execute at most $B_k \in \mathbb{Z}_{\geq 1}$ jobs at once and
costs $c_k \geq 1$ per time unit.

When we say that a machine is running in the interval $[\alpha,\beta]$, we mean that the machine is running at all timeslots in the inclusive interval. For example, in $[\alpha,\alpha]$, the timeslot $\alpha$ is active, and in $[\alpha,\alpha+1]$ timeslots $\alpha$ and $\alpha+1$ are active.
Thus the cost of running a machine of type $k$ for all timeslots in the interval $[\alpha,\beta]$ is $c_k \cdot (\beta-\alpha+1)$. 
This cost is incurred regardless of how many jobs are running on the machine
(but we assume without loss of generality that at least one job is running,
as otherwise we can reduce or split the interval).
Also, jobs can be assigned to a machine that is already processing other jobs, as long as there is spare capacity,
though this may result in the machine running longer.

The labeling on the machine types is such that the capacity of the machines is increasing, 
i.e., $B_0 < B_1 < \ldots$.
Note we assume the ordering is strictly increasing, 
as if there were ever machines with the same capacity but different costs, 
one would always use the cheaper machine of the same capacity.
Without loss of generality,
machines with larger capacity should cost more; e.g.,
if $c_1 \leq c_0$, 
then an optimal schedule could always use type 1 machines instead of type 0 machines. One can assume that there are only finitely many types
as we have finitely many jobs.
After these considerations, let $K$ denote the number of remaining machine types. Recall that the number of machines of each type is unlimited.

For unit-processing-time jobs,
a \emph{batch} of jobs $X$ is defined by 
three features: 
(1) $J(X)$, the set of jobs the batch contains, 
(2) $t(X)$, the batch's type, which corresponds to the machine type that the jobs in $J(X)$ should be executed on, and
(3) $\tau(X)$, the time at which the jobs in $J(X)$ are executed on a machine of type $t(X)$.
We denote a family of batches by $\mathcal{X}$.
A feasible schedule is a partition of jobs into batches such that
every batch $X$ has $|J(X)| \leq B_{t(X)}$,
and every $j \in J(X)$ has $r_j \leq \tau(X) \leq d_j$.

An optimal offline solution is denoted by \textsf{OPT}, 
with cost denoted by cost(\textsf{OPT}).
The time horizon is denoted by $[0,T]$,
with timeslots $\tau \in \{0, 1, \ldots, T\}$, where
sometimes timeslots have subscripts or take an argument.
Continuous intervals with interval endpoints in $[0,T]$ are denoted by $I$, also sometimes with subscripts or taking an argument.

\section{Warm-up: Agreeable Deadlines, Unit-processing-time Jobs}\label{sec: agreeable}

In Subsection \ref{sec: agreeable-proof-main}, we present a 2-competitive algorithm for the setting when unit-processing-time jobs have agreeable deadlines.
In Subsection \ref{sec: adversary-2}, we show that this result is essentially best possible.
Then in Subsection \ref{sec: tech-overview}, we discuss extending algorithms to general job deadlines, with unit-processing-time jobs.

\subsection{Proof of Theorem \ref{thm: agreeable}}\label{sec: agreeable-proof-main}

We will show that Algorithm \ref{algo:greedy} is 2-competitive when jobs have agreeable deadlines. 
Algorithm \ref{algo:greedy} is a straightforward greedy algorithm---it collects jobs until a deadline is reached,
and then it creates optimal batches for scheduling the set of
all the waiting jobs. We use $W$ to represent the set of waiting jobs.

\begin{algorithm}[htb!]
\caption{Greedy}
\label{algo:greedy}
\begin{algorithmic}
\State Let $W \leftarrow \emptyset, \mathcal{X} \leftarrow \emptyset$.
\For{$\tau \leftarrow 0$ to $T$}
     \State $\rhd$ Let $J'$ be the set of jobs with release time $\tau$.
    \State $\rhd$ Update $W \leftarrow W \cup J'$.
    \If{$\exists$ $j^* \in W $ with $d_{j^*} = \tau$}
        \State $\rhd$  $\mathcal{X}_{\tau} \gets \GSS(W)$.
        \State $\rhd$  Execute all batches in $\mathcal{X}_{\tau}$. {\emph{ // Thus completing all jobs in $W$}}
        \State $\rhd$ Update $\mathcal{X} \leftarrow \mathcal{X} \cup \mathcal{X}_{\tau}$.
        \State $\rhd$ $W \leftarrow \emptyset$.
    \EndIf
\EndFor \\
\Return the schedule $\mathcal{X}$.
{\emph{ // With cost at most 2 $\cdot$cost(\textsf{OPT})}}
\end{algorithmic}
\end{algorithm}

The subroutine $\GSS(A)$ takes as input a set of jobs $A$. 
It first finds a minimum cost multiset of machines with total capacity at least $|A|$,
then fills these machines with the jobs in $A$, outputting a set of new batches.
It can be implemented in time $O(n \cdot K)$ using a classic dynamic program, 
where recall that $K$ is the number of distinct machine types.
We can initialize the optimal cost of scheduling $0$ jobs to $0$.
After initialization, we can then optimally compute the cost of scheduling
$|A|$ jobs by solving the following recurrence:
\[
   \text{cost}(|A|) = \min_{k \in \{0,\ldots,K-1\}} \: \text{cost}(|A| - \min(B_k, |A|)) + c_k.
\]

In the offline setting,
this computation can be made once before Algorithm
\ref{algo:greedy} starts,
and the optimal values stored in a table that can be used whenever
Algorithm \ref{algo:greedy} needs them. 
In the online setting, we do not know $n$ in advance, so
the table is extended lazily whenever the largest waiting-set 
size increases.  The time spent on such an extension is $O(K)$ times
the number of  new jobs, and the total time used for
increasing the table size is $O(n)$ (by the classical ``double-the-size" method as in \cite{CLRS09}).

Offline,
we sort $J$ by release times so that $J'$ can be found in time $O(|J'|)$. This involves simply maintaining an ordered set of jobs.
Online,  arrivals are already revealed chronologically.
To store $W$, we use a min-heap with keys $d_j$, for elements $j \in W$.
We do not have to go timeslot by timeslot, but instead use the sorted
$J$ to find the next time where $J' \neq \emptyset$ and compare
this with the smallest key in $W$.
The overall running time is $O(n K + n \log n) $.

The following lemma crucially (though perhaps subtly) uses the agreeable deadlines structure.
As in Algorithm \ref{algo:greedy}, let $\mathcal{X}_\tau$ be the set of batches executed at time $\tau$.
    For each set of batches $\mathcal{X}_\tau$,
    construct an interval $I_{\tau}$ starting at the earliest
    release time of any job scheduled in a batch in $\mathcal{X}_\tau$
    and ending at the latest deadline of any job in a batch in 
    $\mathcal{X}_\tau$.

\begin{lemma}\label{two_disjoint}
    At most two of the intervals $I_{\tau}$ as defined above overlap at any point in time.
\end{lemma}
\begin{proof}
    It suffices to prove that each such interval contains at most one left 
    endpoint of another interval. 
From the way the algorithm schedules all the jobs in $W$,
we obtain that for any $\tau$ there exists at most one
$\mathcal{X}_{\tau}$. Assume we have
$\mathcal{X}_{\tau}$ and $\mathcal{X}_{\tau'}$ with $\tau < \tau'$.
Then the left endpoint of $I_{\tau'}$ is strictly to the right of $\tau$,
as any job released at or before $\tau$ that has not been
scheduled  before $\tau$ is scheduled in $\mathcal{X}_{\tau}$. 
This also implies that the left endpoint of $I_{\tau'}$ is strictly
to the right of the left endpoint of $I_{\tau}$.
    Call two such intervals $I_{\tau}, I_{\tau'}$
    \textit{consecutive} if $\tau < \tau'$ and there is no
   $\tau''$ with $\tau < \tau'' < \tau'$  such that the algorithm produces an
$\mathcal{X}_{\tau''}$.  If $I_{\tau}, I_{\tau'}$ are consecutive,
then there is no $I_{\tau''}$ whose left endpoint is between the
left endpoints of   $I_{\tau}$ and $I_{\tau'}$.

    Consider three consecutive intervals $I_{\tau_1}, I_{\tau_2}, I_{\tau_3}$.
    For interval $I_{\tau_1}$, let job $j_1$ be a job
    scheduled in $\mathcal{X}_{\tau_1}$  with the earliest release time,
    breaking ties in favor of the job with the earliest deadline.
    Also for $I_{\tau_1}$, let job $j_1'$ be a job  scheduled in 
    $\mathcal{X}_{\tau_1}$ 
    with the latest deadline, breaking ties arbitrarily.
    Let $j_2, j_2', j_3$, and $j_3'$ be defined analogously for 
    $I_{\tau_2}$ and $I_{\tau_3}$.

    In order to show that $I_{\tau_1}$ and $I_{\tau_3}$ do not overlap, we will show that 
    \begin{equation}\label{eqn: agreeable}
        d_{j_1'} \leq d_{j_2} = \tau_2 < r_{j_3}.
    \end{equation}

    We first argue that $\tau_2 = d_{j_2}$.
    Indeed, we must have $\tau_2 \leq d_{j_2}$ since $j_2$
    is scheduled in $\mathcal{X}_{\tau_2}$.
    On the other hand, if it were the case that $\tau_2 < d_{j_2}$, 
    then there is some job $j$  scheduled in $\mathcal{X}_{\tau_2}$
    with deadline $d_j = \tau_2$, 
    and jobs $j$ and $j_2$ break the agreeable relation because
    $d_j < d_{j_2}$  and $r_j > r_{j_2}$ (by the way $j_2$ is selected).
    See Figure \ref{f:deadline}.

\begin{figure}[th]
\begin{center}\leavevmode%
\scalebox{0.25}{
  \includegraphics{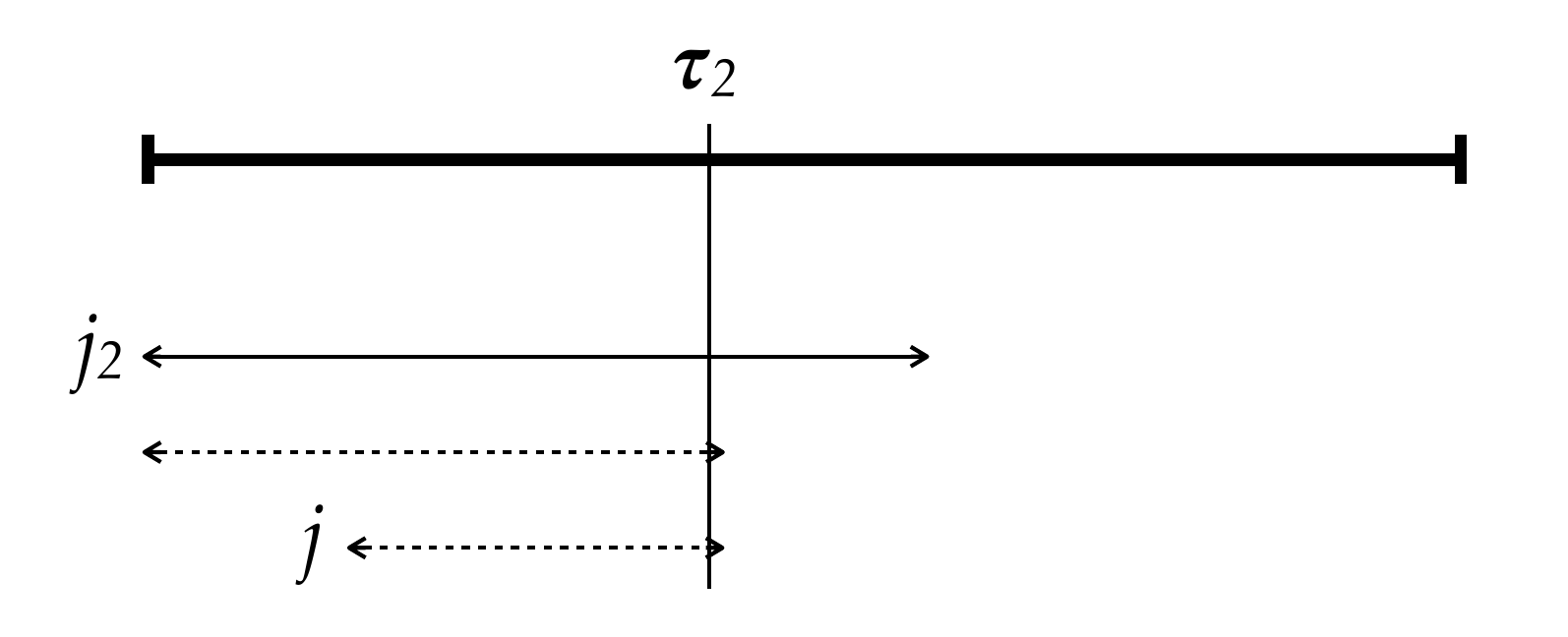}
}
\end{center}
\caption{An illustration of an argument
from the proof of Lemma \ref{two_disjoint}, precisely why $d_{j_2} = \tau_2$:
If $\tau_2 < d_{j_2}$, then one of the dashed jobs/intervals
must end at $\tau_2$. However, the upper dashed job cannot exist since $j_2$
is chosen as having the earliest deadline among the jobs scheduled in
$\mathcal{X}_{\tau_2}$ that have the earliest release time,
while the lower dashed job cannot exist since
it is not agreeable with $j_2$.
}
\label{f:deadline}
\end{figure}

    The last inequality of Equation (\ref{eqn: agreeable}) follows
    from the discussion in the first paragraph of the proof, 
    with $\tau_2$ instead of $\tau$ and $\tau_3$ instead of $\tau'$. The same reasoning gives that $\tau_1 < r_{j_2}$.
    We have that $r_{j_1'} \leq \tau_1$ since $j_1'$ is scheduled in 
    $\mathcal{X}_{\tau_1}$. Thus $r_{j_1'} < r_{j_2}$.
    If $d_{j_1'} > d_{j_2}$, then $j_1'$ and $j_2$ are not agreeable.
    Thus the first inequality of Equation (\ref{eqn: agreeable}) also holds.

    From the inequalities in (\ref{eqn: agreeable}), the left endpoint of $I_{\tau_3}$ is not in $I_{\tau_1}$.
    For $\tau' < \tau_1$, we have seen before that the left endpoint of $I_{\tau'}$ is strictly before
    the left endpoint of $I_{\tau_1}$, and therefore not in $I_{\tau_1}$.
    For $\tau' > \tau_3$, as we have seen before,
    the left endpoint of $I_{\tau'}$ is strictly after  the left endpoint of $I_{\tau_3}$, and thus also
    not in $I_{\tau_1}$. Thus $I_{\tau_1}$ indeed can contain at most one left endpoint of another interval
    (namely $I_{\tau_2})$, completing the proof.
\end{proof}

If deadlines are not agreeable, 
many such intervals can overlap at any time point. 
In the next section, we use a more complicated interval construction to 
lower bound cost(\textsf{OPT}).

\smallskip

\begin{proof}[\emph{Theorem \ref{thm: agreeable}}]
As before, 
construct an interval $I_{\tau}$
for each set of batches $\mathcal{X}_\tau$ dispatched
by the Greedy algorithm,
where the left endpoint of the interval corresponding to $\mathcal{X}_\tau$ is the earliest release time of any job in $ \cup_{X \in \mathcal{X}_\tau}J(X)$ and the right endpoint is the latest deadline of any job in $\cup_{X \in \mathcal{X}_\tau}J(X)$.
Lemma \ref{two_disjoint} implies that we can partition the intervals into two sets $S_1$ and $S_2$ of disjoint intervals.
Specifically, for any two consecutive, overlapping intervals, it suffices to have one such interval in $S_1$ and the other in $S_2$.
Let $J_1$ be the set of jobs executed by batches in $\mathcal{X}_\tau$
with $I_{\tau} \in S_1$, and $J_2$ likewise for $S_2$.
Let $\textsf{OPT}(J_i)$ denote an optimal offline schedule for $J_i$, for $i=1,2$.
Note that for any sets of jobs $J$ and $J'$,  if $J' \subseteq J$,
then $\text{cost}(\textsf{OPT}(J')) \leq \text{cost}(\textsf{OPT}(J))$. Then:
\begin{align*}
    \text{cost(Greedy)} &= \text{cost}(\textsf{OPT}(J_1)) + \text{cost}(\textsf{OPT}(J_2))
    \leq 2 \cdot \text{cost(\textsf{OPT})}
\end{align*}
The equality is because the intervals in $S_1$ are disjoint and the intervals in $S_2$ are disjoint, and we have scheduled all jobs in each interval using the minimum possible cost with $\GSS$.
\end{proof}

\subsection{Proof of Theorem \ref{thm: lb}}\label{sec: adversary-2}

Here, we prove a lower bound of 2 for the competitive ratio of any deterministic online algorithm, assuming an all-powerful
(also called \emph{adaptive-offline} \cite{BEY98}) adversary. 
Our construction results in an agreeable instance of unit-processing-time jobs,
which implies that Algorithm \ref{algo:greedy} is an optimal deterministic
online algorithm for agreeable instances. 

\medskip

\begin{proof}[\emph{Theorem \ref{thm: lb}}]
Suppose that there are two machine types, 
which we call ``small" and ``large". 
The small machines have cost $1$ and capacity $1$,
whereas the large machines have cost $M$ and capacity $M^3$, 
where $M$ is a large even integer. 
We call batches run on these types large and small batches, respectively.
Here, jobs have unit processing time.
In this instance, the release time of every job will be odd and the deadline of every job will be even. 
Many jobs will have the same release time, 
but no two jobs will have the same deadline.
The adversary's only action will be to release groups of $(M^3)/2$ jobs, 
where those $(M^3)/2$ jobs arrive at some time $\tau$ and each released job has as deadline the earliest even timeslot that does not already contain a job deadline (thereby adding $(M^3)/2$ unique deadlines to the instance). 
Note that this release structure will result in an agreeable instance.

At time $1$, the adversary releases one group of $(M^3)/2$ jobs. 
If no large machine is used to execute any of the jobs from the current group, 
then no more jobs are released by the adversary. 
Otherwise, the adversary will release another group of $(M^3)/2$ jobs at the first odd timeslot after the algorithm uses a large machine.
The adversary repeats this at most $M - 1$ more times, so that we have at most $M$ such groups. We index the groups of jobs released by the adversary as $\beta_1,\ldots, \beta_g$.

\medskip

We now show why this instance implies a lower bound of 2 on the competitive ratio. 
First, consider the case where the algorithm uses at least $2M^2$ small machines in a row in some group. These batches would incur a cost of at least $2M^2$. One 
feasible schedule across all groups of jobs would be to have instead 
executed all of the jobs from each group in a large batch at the first deadline of a job from each group, which would have incurred a cost of at most $M^2$ total. 
This implies a competitive ratio of at least $2$. 

\begin{figure}
    \centering
    \includegraphics[width = 13cm]{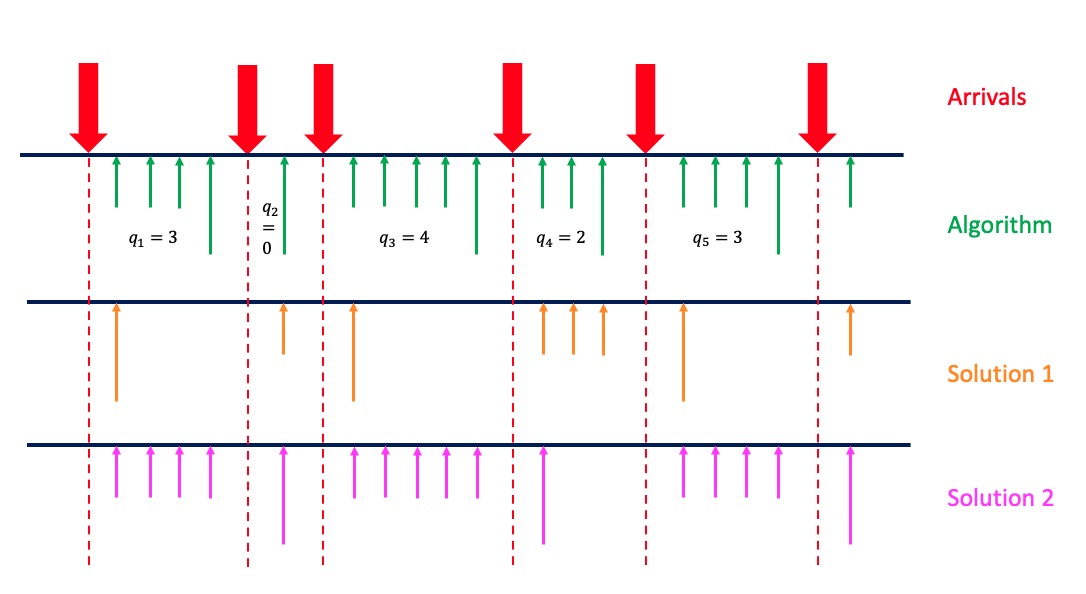}
    \caption{An illustration for the proof of Theorem \ref{thm: lb}.
    Each red thick arrow represents the arrival of $(M^3)/2$  jobs. 
    The short thin arrows represent the scheduling of a batch on a small machine, and the long thin arrows 
    represent the scheduling of a batch on a large machine.
    }
    \label{f_lb_2}
\end{figure}

See Figure \ref{f_lb_2} for an illustration of the rest of the proof.
If an algorithm does not schedule
$2M^2$ small batches in a row in any group,
then it executes at least one large batch for every group. 
Note that in this case all $M$ groups are released by the adversary,
and $g=M$.
Let $q_i$ represent the number of small batches scheduled by the algorithm between the releases of groups
$\beta_{i}$ and $\beta_{i + 1}$ for $i \in [M-1]$, 
and let $q_M$ be the number of small batches scheduled by the algorithm after $\beta_M$ was released. 
Then the cost incurred by the algorithm is 
at least $M^2 + \sum_{i = 1}^M q_i$. 

We consider two other possible solutions to this instance (see Solution 1 and Solution 2 in Figure \ref{f_lb_2}). 
The first possible solution is to schedule a large batch at the first deadline, 
then small batches at the next $q_2 + 1$ deadlines, 
then a large batch at the next deadline, 
then small batches at the next $q_4 + 1$ deadlines, 
and so on, ending with a large batch to clear away any remaining jobs. 
This solution is feasible as we can map any job executed by the algorithm to a batch in this first solution,
as follows. Any job executed in a small batch in an even group in the algorithm is also executed by 
a small batch in this first solution. 
The job with the earliest deadline that is in a large batch of the algorithm in every even group 
is also in a small batch of this first solution. Every other job executed by the algorithm is in a large batch
in this first solution.
Since the algorithm scheduled all jobs, we know this first solution scheduled all jobs as well. 
The cost of this first solution is at most 
$(M/2 + 1) M +  \sum_{i=1}^{M/2}  (1 + q_{2i}) = (M + 3)(M/2) +  \sum_{i = 1}^{M/2} q_{2i}$. 

A second possible solution is to use small batches at the first $q_1 + 1$  deadlines,
then a large batch at the next deadline, then small batches at the next $q_3 + 1$ deadlines,
and so on, ending with a large batch to clear away any remaining jobs. 
We can again show this second solution is feasible by mapping jobs scheduled
by the algorithm to jobs
scheduled by this solution in a similar fashion. The cost of this second solution is at most 
 $(M/2 + 1) M +  \sum_{i=1}^{M/2}  (1 + q_{2i-1}) = (M + 3)(M/2) +  \sum_{i = 1}^{M/2} q_{2i-1}$. 
The sum of the costs for these two possible solutions is at most $(1 + 3/M)$ times the cost of the online algorithm. This then implies that the solution with the lower cost of the two is at most $(1 + 3/M)/2$ times the cost of the online algorithm.
Letting $M$ go to infinity completes the proof of a lower bound of 2 for the competitive ratio of any deterministic online algorithm in the all-powerful adversary model.
\end{proof}

\subsection{Difficulty in non-agreeable deadlines}\label{sec: tech-overview}
Any algorithm for this problem with unit-processing-time jobs
can execute batches only when it is forced to, i.e., when the current timeslot $\tau$ is the deadline of some uncompleted job $j^*$, as there is nothing to be gained from executing a batch before a deadline. 
Additionally, jobs can always be assigned to an open batch
by an earliest deadline first (EDF) rule---that is, as long as a batch has space,
repeatedly select from the set of jobs released and not yet assigned to a batch 
the job with the earliest deadline, and assign it to the batch.
It follows that the only remaining decision for an algorithm is what type of batch (and hence machine) should be used to execute job $j^*$ at time $\tau$; 
note that another way to phrase this question is how many waiting jobs should be processed with $j^*$.
An adversary can punish an algorithm for processing too many waiting jobs at a timeslot by dispatching a huge group of jobs soon after (as it would have been more cost-efficient to execute every job together on a machine with large capacity).
On the flip side, the adversary can punish an algorithm for processing small batches by not sending any new jobs (as a large batch would have been more cost-efficient here).
We discuss why some simple heuristics to choose the machine type fail in Appendix \ref{sec: app-algs}. 

Agreeable deadlines were easy to handle because we can charge the cost of
the Greedy algorithm to simple lower bounds on cost(\textsf{OPT}), 
and recall that the lower bounds were due to the nice structure of the intervals corresponding to the batches created by Greedy.
For general deadlines, we are able to track a more
complicated set of intervals corresponding to batches in order to decide  
which machine type to use at the deadline of an uncompleted job.
Moreover, this complicated interval structure implies we need a more careful lower bound on cost(\textsf{OPT}).
Specifically, our algorithm tracks a set of nested intervals
$[r_{j^*},\tau]=I_0 \subseteq I_1\subseteq \cdots \subseteq I_k$,
all with right endpoint $\tau$, where for every 
$\ell \in \{0,...,k-1\}$, 
a batch of type $\ell$ was already executed in interval $I_\ell$. Eventually, our method of constructing this sequence of nested intervals produces an interval $I_{\ell^*}$ that contains no timeslot where a batch of type $\ell^*$ was executed. This indicates that our algorithm should create a batch of type $\ell^*$ at time $\tau$.

\section{Online Algorithms for the General Case}\label{sec: lbp+proof}

The algorithms of Theorems \ref{thm: deterministic} and \ref{thm: randomized} are based on the core algorithm from the following lemma:

\begin{lemma} \label{lem: main}
    Assume all the costs are powers of $2$ and all machines have to be used for exactly $2p-1$ timeslots 
    (even if no jobs are scheduled at some of these timeslots).
    Then there exists a deterministic $4$-competitive online algorithm for 
    Uniform-processing-time Busy Time Scheduling on Heterogeneous Machines.
    The algorithm's running time is $O(n \log n)$, for $n$ the total number of jobs.
\end{lemma}

We prove a lower bound on cost(\textsf{OPT}) in Subsection \ref{sec: plower-bound}, 
and then use it to design and analyze our algorithm that proves Lemma \ref{lem: main} in Subsection \ref{sec: palgs}.
Let us assume this lemma for the moment and proceed to the algorithms for the general case
(Theorems \ref{thm: deterministic} and \ref{thm: randomized}).

\subsection{Preliminaries to prove Theorems \ref{thm: deterministic} and  \ref{thm: randomized} assuming Lemma \ref{lem: main}}\label{sec: extra-prelim}

We first show that we can assume that all machines run for exactly $2p-1$ 
timeslots, and only lose a factor of $(2p-1)/p$ in the competitive ratio. Intuitively, it is helpful to think about machines running for $2p-1$ timeslots
because any two jobs which are batched together at any point are scheduled in
an interval of length at most $2p-1$ timeslots.

\begin{proposition}\label{prop: 2p-1}
    When all jobs have uniform processing time 
    $p$, we can assume that  machines run for exactly $2p-1$ timeslots 
    (thus each batch has $2p-1$ timeslots),
    while only losing a factor of $(2p-1)/p$ in the competitive ratio.
\end{proposition}

\begin{proof}
Suppose that a feasible solution runs a machine of type $k$ at all timeslots in the interval
$[\alpha,\beta]$,  for a cost of $c_k \cdot ( \beta - \alpha + 1)$.
Note it must be that $\beta \geq \alpha + p -1$, or else no job could be scheduled
on this machine. Let $\gamma = \lfloor (\beta - \alpha + 1)/p \rfloor$.
Another feasible schedule runs $\gamma$ machines of type $k$ for the intervals 
$[\alpha + i \cdot p, \alpha + (i+2) p -2]$,
for each $i = 0, \ldots, \gamma-1$.
This solution has a total cost of $c_k\cdot  \gamma \cdot (2p-1)$.
One can easily check that $\gamma \cdot (2p-1) \leq \frac{2p-1}{p} (\beta - \alpha + 1)$
and that every job scheduled on the original machine fits on 
exactly one of the new machines. Moreover, at no time is a new
machine over capacity, as otherwise the original schedule must have been over capacity at the 
midpoint of the processing time of that new machine.
\end{proof}

We now use a standard bucketing technique.
We show in the following proposition that
we can assume that all machine costs are powers of 2,
while losing only a factor of 2 in the competitive ratio.

\begin{proposition}
\label{prop: pcost-ratios}
At a loss of at most a factor of 2 in the competitive ratio,
we can assume
that for all $k \in \mathbb{Z}_{\geq 0}$, $c_k = 2^q$ for
some $q \in \mathbb{Z}_{\geq 0}$, and $c_0=1$.
\end{proposition}
\begin{proof}
First, if $c_0 \neq 1$, scale all of the machine costs by $1/c_0$.
Note this does not
affect the optimality of any schedules.
For each $k$, let $q_k$ be the smallest integer such that $c_k \leq 2^{q_k}$.
Then take $c_k' = 2^{q_k}$ to be the new machine cost for each $k$.
Note that the cost of
a schedule $\mathcal{S}$ with machine costs $c'$ is at most 2 times
the cost of schedule $\mathcal{S}$ with machine costs $c$.
\end{proof}

\begin{proof}[\emph{Theorem \ref{thm: deterministic}}]
    The statement follows immediately from
Lemma \ref{lem: main}, Proposition \ref{prop: 2p-1}, and Proposition \ref{prop: pcost-ratios}.
\end{proof}

Using randomization, we can obtain a better competitive ratio by replacing the
proposition above by the next one (a variant of this was used by \cite{SLLA25}, 
and we first saw this technique in the 2000 conference version of \cite{CNS04}).

\begin{proposition}
\label{prop: pcost-ratios-rand}
At a loss of at most a factor
of $1/\ln 2$ in the competitive ratio,
we can assume
that for all $k \in \mathbb{Z}_{\geq 0}$, $c_k = 2^q$ for
some $q \in \mathbb{Z}_{\geq 0}$, and $c_0=1$.
\end{proposition}
\begin{proof}
    Randomly and uniformly choose $y$ in the range $[0,1]$ and replace each $c_l$ by $c'_l = 2^{i+y}$,
where $i$ is the smallest integer satisfying $2^{i+y} \geq c_l$.
Let $x$ be such that $c_l = 2^{x+j}$, where $j$ is an integer and $x \in [0,1)$.
If $x \leq y$, then $c'_l = 2^{y-x} c_l$, and when $x > y$, then $c'_l = 2^{y+1-x} c_l$.
Since $y$ is chosen uniformly in the range $[0,1]$, we get that the expected value of $c'_l$ is $c_l$ times
\[
    \int_{0}^{x}  2^{y+1-x} dy + \int_{x}^{1} 2^{y-x} dy = \frac{2 - 2^{1-x}}{\ln 2} + \frac{2^{1-x} -  1}{\ln 2} = \frac{1}{\ln 2}.
\]
Consider an optimum solution with the original costs $c_k$ and let $\opt$ be its cost.
Let $\cost'(\OPT_c)$ be the cost of this solution with modified costs $c'_k$.
By linearity of expectation, 
\begin{equation*}
    \mathbb{E}[\cost'(\OPT_c)] \leq \frac{1}{\ln 2} \cdot  \opt.
\end{equation*}

Lastly, divide all $c'_k$ by $c'_0$, so all costs are powers of $2$ and the smallest cost is 1.
\end{proof}

\begin{proof}[\emph{Theorem \ref{thm: randomized}}]
The statement follows from Lemma \ref{lem: main}, Proposition \ref{prop: 2p-1}, and  Proposition \ref{prop: pcost-ratios-rand}. 
\end{proof}

\subsection{Preliminaries to prove Lemma \ref{lem: main}}\label{sec: further}

In the context of proving Lemma \ref{lem: main}, for simplicity, we drop the factor of $2p-1$ from the objective.
 \begin{assumption}\label{assm: 2p-1}
For a machine of type $k$, we only count $c_k$ in the objective function,
ignoring a multiplicative factor of $(2p-1)$.
\end{assumption}

For uniform-processing-time jobs,
a \emph{batch} of jobs $X$ is defined by
three features:
(1) $J(X)$, the set of jobs the batch contains,
(2) $t(X)$, the batch's type, which corresponds to the machine type that the jobs in $J(X)$ should be executed on, and
(3) $\tau(X)$, the  start time.
This is a generalization of the definition from Subsection \ref{sec:prelims}
where unit-processing-time jobs were considered.
For batch $X$, the machine of type $t(X)$ is on for timeslots $\tau(X), \tau(X) + 1, \ldots, \tau(X) + 2p -2$
(a total of $2p-1$ timeslots, as in the statement of Lemma \ref{lem: main}).
The job $j \in J(X)$, if $d_j \leq \tau(X) + 2p -2$, is executed in timeslots  $d_j - p +1, \ldots, d_j$.
If $d_j > \tau(X) + 2p -2$, then the job $j$ is  executed in timeslots  $\tau(X) + p -1, \ldots, \tau(X) + 2p -2$.
For batch $X$ to be feasible, the timeslots we assigned above to job $j$  must all be in the interval $[r_j,d_j]$,
and  $|J(X)| \leq B_{t(X)}$ 
(this suffices since all the jobs in $J(X)$ are scheduled to be executed in timeslot $\tau(X) + p -1$).
We denote a family of batches by $\mathcal{X}$.

Without loss of generality, further assumptions can be made.

\begin{assumption}\label{assm: pcost-ratios}
For $k \in \mathbb{Z}_{\geq 0}$, a machine of type $k$ has cost $c_k = 2^k$.
\end{assumption}
Indeed, we can scale all the costs by the same factor so that the machine of lowest cost has cost $2^0 = 1$. 
Recall that (in Lemma \ref{lem: main}) all machine costs are powers of 2.
If a machine of cost $2^{k}$ is not present for some $k \in \mathbb{Z}_{\geq 0}$,
one could just use $2^{k-k'}$ machines of cost $2^{k'}$, for  $k'$
the largest type less than $k$ that is present.

This assumption is useful in discussing the cost-per-job of the different machine types, as seen in the following proposition.

\begin{proposition}\label{prop: pcost-per-person}
Given Assumption \ref{assm: pcost-ratios}, 
one can also assume without loss of generality  that
the cost-per-job is non-increasing in the machine types,
i.e., $c_0 / B_0 \geq c_1 / B_1 \geq \ldots $.
\end{proposition}
\begin{proof}
Suppose we are given machine types satisfying 
Assumption \ref{assm: pcost-ratios}.
If ever there existed $k$ with
$ c_{k}/ B_{k} <  c_{k+1} / B_{k+1}$, 
then there always exists an optimal schedule 
that does not use any machines of type $k+1$. 
This is because any schedule that used machines of type $k+1$
can replace each machine of type $k+1$
with 2 machines of type $k$, since by Assumption \ref{assm: pcost-ratios} machine costs are powers of 2.
\end{proof}

Note: we used here that $c_k = 2^k$. If $c_k = \rho^k$ for
some non-integer $\rho$, we are not able to get
$c_i/B_i  \geq  c_{i+1}/B_{i+1}$ above, 
and this inequality is used later in Lemma \ref{lem: pvalid-assgn-lb}.
Assumption \ref{assm: pcost-ratios} and Proposition \ref{prop: pcost-per-person} also allow us to upper-bound  $K$, the number of distinct machine types.

\begin{proposition}\label{prop:pUB-K}
Given Assumption \ref{assm: pcost-ratios}, 
the number of distinct machine types is bounded,
with $K \leq 2 + \log_2 n$.
\end{proposition}
\begin{proof}
    By Assumption \ref{assm: pcost-ratios}, a machine of type $k$ has cost $c_k=2^k$, and by Proposition \ref{prop: pcost-per-person}, $c_k/B_k \geq c_{k+1}/B_{k+1}$. Substituting in for the costs and rearranging, we see that $B_{k+1}\geq 2 B_k$. 
    The machine capacities are geometrically increasing by at least a factor of 2.
    Since there are at most $n$ jobs,
    the highest capacity machine that is relevant to the instance processes a batch of size at most $2n-1$. It follows that $K \leq 2 + \log_2 n$.
\end{proof}

It will be helpful to track intervals $I \subseteq [0,T]$, 
and often we associate a type and a set of jobs with interval $I$.
We let $t_I$ denote the associated type and $J_I \subseteq J$ the associated set of jobs. 

Consider an instance of the variant of Flexible Busy Time Scheduling on Heterogeneous Machines  discussed in Lemma \ref{lem: main}, and also 
the instance satisfies Assumption \ref{assm: pcost-ratios}.
Let $\hOPT$ be its optimum solution, and let $\cost(\hOPT)$ be its cost when we ignore factors of $2p-1$, as in Assumption  \ref{assm: 2p-1}.

\subsection{A lower bound on $\cost(\hOPT)$}
\label{sec: plower-bound}

\begin{figure}
    \centering
    \includegraphics[width =13cm]{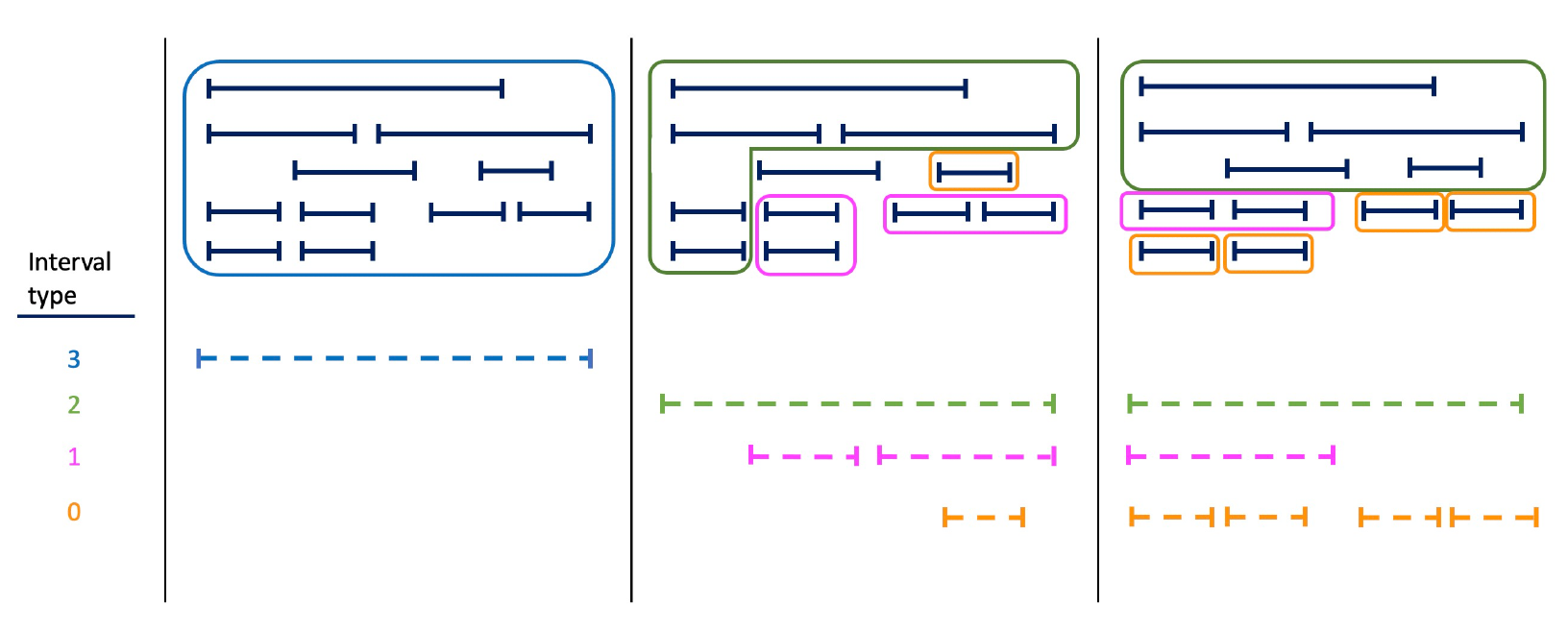}
    \caption{There are 11 unit-processing-time jobs, each of which is depicted by a solid black line based on its release time and deadline. The three different frames  depict possible valid interval assignments for the jobs. The dashed lines of different colors represent intervals of different types in each frame. There are four types of machines, with costs $2^k$ for $k \in \{0, 1, 2,3\}$ and with capacities $\{2, 5, 11, 22\}$. Each interval assignment implies a different lower bound for $\cost(\hOPT)$, as in Lemma \ref{lem: pvalid-assgn-lb}. We obtain better lower bounds moving from left-to-right in the frames, with bounds $2, 9/4$, and $5/2$, respectively. 
    Note that an optimal schedule here has cost 5, with 3 batches of type 0 and 1 batch of type 1.}
    \label{fig:intervals-LB}
\end{figure}

We consider the following definition, for which Figure \ref{fig:intervals-LB} is an accompaniment.
\begin{definition}\label{def: pcharacterization}
For a set of jobs $J$ equipped with 
release times and deadlines in $[0,T]$, 
let an \emph{interval assignment} $\mathcal{A}$ be a family of tuples $\{(I,t_I, J_I)\} $ of a continuous interval $I \subseteq [0,T]$, a type $t_I$, and a set of jobs $J_I \subseteq J$. 
Let $\cL$ denote the multiset of all intervals $I$ with $(I, t_I, J_I) \in \mathcal{A}$.
We define a \emph{valid interval assignment} as one such that the following holds:
\begin{enumerate}
\item  When $t_I \geq 1$, then $|J_I| = B_{t_I - 1}$, and when $t_I=0$, then $|J_I| = 1$.
\item If job $j$ is in $J_I$, then $[r_j,d_j] \subseteq I$.
\item For any two distinct tuples in $\mathcal{A}$ and their intervals
$I, I'$, if $t_I = t_{I'}$, then $I$ and $I'$ are disjoint.
\item Every job is assigned to at most one $J_I$. 
\end{enumerate}
\end{definition}

Let $I(j)$ be the interval to which job $j$ is assigned. 
If a job $j$ is not assigned to any $J_I$, then  we write $I(j) = \emptyset$.

In other words,
a valid interval assignment represents a partial partition of jobs such that each job is assigned to at most one interval which fully contains the job's feasible region. Each interval is assigned a type such that no two intervals of the same type overlap. Furthermore, an interval 
is assigned exactly the number of jobs specified in
the first condition above. 
Intuitively, valid interval assignments are designed to
prevent an algorithm from ramping up too fast when ``disjoint"
batches of the same type are possible.
We use valid interval assignments to lower bound the cost of an optimal offline solution, as in the following lemma.

\begin{lemma}\label{lem: pvalid-assgn-lb}
    Consider an instance  satisfying Assumption \ref{assm: pcost-ratios}
    and Proposition \ref{prop: pcost-per-person}, and
    let $\mathcal{A}=\{(I,t_I, J_I)\}$ be a valid interval assignment with $\cL$ the multiset of all intervals $I$ with $(I, t_I, J_I) \in \mathcal{A}$. Then,
    \[
        \cost(\hOPT) \geq \frac{1}{4} \cdot \sum_{I \in \cL} 2^{t_I}.
    \]
\end{lemma}
\begin{proof}   
    At a high level, we will prove this by  introducing  at most
    $4 \cdot \cost(\hOPT)$ credits and then distributing them
    to all jobs $j$ for which $I(j) \neq \emptyset$. We will do this in such a way that the sum of the credits distributed can be shown to be
    at least $\sum_{I \in \cL} 2^{t_I}$,
    proving the inequality in the lemma statement.

    Suppose that for each batch $X$ of type $k$ that $\hOPT$ uses, we distribute credits to jobs in $J(X)$ in the following way:
    \begin{itemize}
        \item To each job $j$ in $J(X)$ with $t_{I(j)} \leq k$ we give $2^{t_{I(j)}}/|J_{I(j)}|$ credits.
        \item To each job $j$  in $J(X)$ with $t_{I(j)} > k$, we give $2^{k + 1}/B_k$ credits.
    \end{itemize}

    Note that, if $I(j) = \emptyset$, then $j$ does not receive any credits.
    Figure \ref{fig:credits} depicts the distribution of credits.

    \begin{figure}
        \centering
        \includegraphics[width = 13cm]{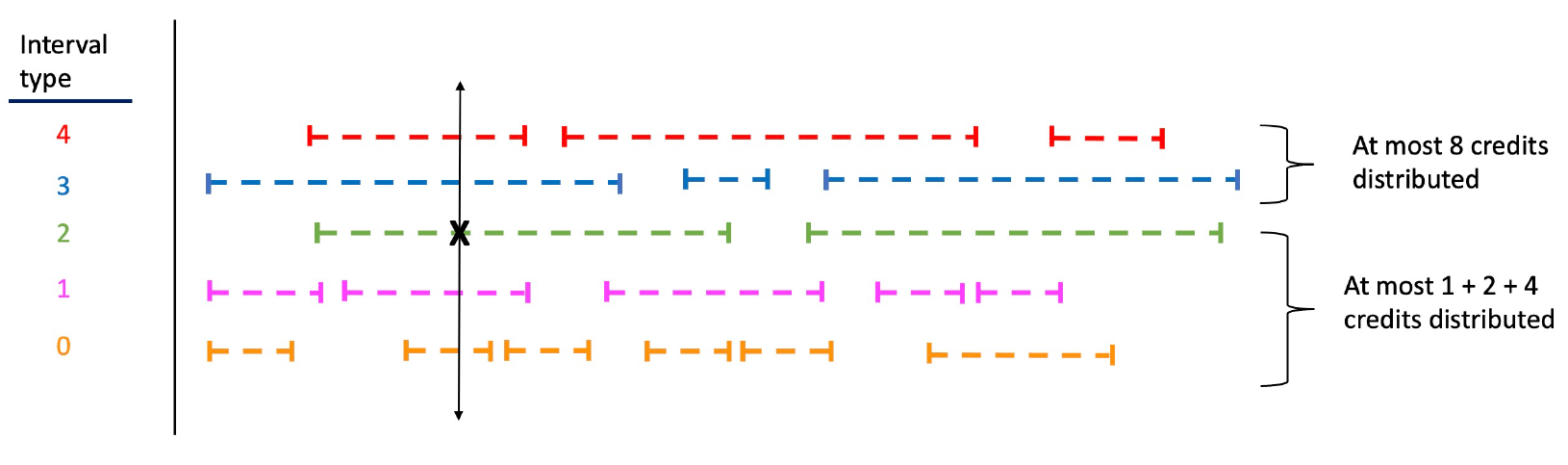}
        \caption{Distribution of credits from batch $X$ of type $k = 2$. The vertical line represents the midpoint of the batch, i.e. $\tau(X) + p -1$.
        The dashed lines of different colors represent intervals of different types in a valid interval assignment.
        For interval types 0, 1, and 2,
        $X$ distributes $2^{t_{I(j)}}/|J_{I(j)}|$ credits per job, and there are at most $|J_{I(j)}|$ jobs in each interval type. For each job in interval types 3 and 4, $X$ distributes $2^{k+1}/B_k$ credits, and there are at most $B_k$ such jobs in interval types 3 and 4 combined.}
        \label{fig:credits}
    \end{figure}

    We will first prove the following helpful claims regarding credits.
    
    \begin{claim}\label{claim: phelper-1}
        When $t_{I(j)} > k$, the following holds:
        \[
            \frac{2^{t_{I(j)}}}{|J_{I(j)}|} \leq \frac{2^{k + 1}}{B_k}. 
        \]
    \end{claim}
    \begin{proof}
        We observe that:
        \[
            \frac{2^{t_{I(j)}}}{|J_{I(j)}|} = 2 \cdot \frac{2^{t_{I(j)} - 1}}{B_{t_{I(j)} - 1}} \leq 2 \cdot \frac{2^k}{B_k} = \frac{2^{k + 1}}{B_k}
        \]
        where the inequality is because $t_{I(j)} > k$ and we assumed (based on Proposition \ref{prop: pcost-per-person}) that larger machines are more cost-efficient.
    \end{proof}

    \begin{claim}\label{claim: phelper-2}
        The credits distributed by $X$ of type $k$ to jobs with $t_{I(j)} \leq k$ is at most $2^{k + 1}$.
    \end{claim}
    \begin{proof}
        Let batch $X$ use the timeslots in the interval $[\alpha,\beta]$ and let 
        $\tau = \alpha + (p  - 1) = \beta - (p-1)$ 
        (note this equality is from  Proposition \ref{prop: 2p-1}). 
        Any job $j$ in $J(X)$ has $\tau \in [r_j,d_j]$.
        We have required that all intervals with the same type must be disjoint,
        and therefore $\tau$ belongs to at most one interval
        $I$ of each type $k'$  with $k' \leq k$.  
        Moreover, since any job assigned to an interval has
        its release time and
        deadline inside that interval, all the jobs in $J(X)$ that are assigned
        to an interval of type $k'$ are assigned to $I$ above.
        From each such interval $I$, $J(X)$ contains at most $|J_I|$ jobs, and so the total amount of credits awarded by $X$ to jobs associated with $I$ is at most $2^{k'}$. The total amount of credits awarded by $X$ to jobs with $t_{I(j)} \leq k$ is therefore upper bounded by $1 + 2 + \cdots + 2^k \leq 2^{k + 1}$.
    \end{proof}

    \begin{claim}\label{claim: phelper-3}
        The credits distributed by $X$ of type $k$ to jobs with $t_{I(j)} > k$ is at most $2^{k + 1}$.
    \end{claim}
    \begin{proof}
        There are at most $B_k$ jobs in batch $X$, and batch $X$ distributes $2^{k + 1}/B_k$ to each job with $t_{I(j)} > k$, so the total amount of credits awarded to jobs with $t_{I(j)} > k$ is upper bounded by $2^{k + 1}$.
    \end{proof}

    We use these claims to prove the lemma statement. 
    Let $\text{credit}(j)$ denote the amount of credit awarded to 
    job $j$. Then (recall that $|J_I| > 0$):
    
    \[
        \sum_{I \in \mathcal{L}} 2^{t_I} = \sum_{I \in \mathcal{L}} \sum_{j \in J_I} \frac{2^{t_I}}{|J_I|} = \sum_{j: I(j) \neq \emptyset} \frac{2^{t_{I(j)}}}{|J_{I(j)}|} \leq \sum_{j: I(j) \neq \emptyset} \text{credit}(j) \leq 4 \cdot \cost(\hOPT).
    \]
    Note that the first sum is over $\cL$, which is a multiset, 
    so in particular an interval might contribute to the sum multiple times.
    For the first inequality, we note that every job is in $J(X)$ 
    for some batch $X$ that $\hOPT$ uses, and
    therefore every job $j$ associated with an interval $I(j)$
    must receive at least $2^{t_{I(j)}}/|J_{I(j)}|$ credits 
    (by construction and by Claim \ref{claim: phelper-1}). 
    
    For the second inequality, we combine Claims \ref{claim: phelper-2} and \ref {claim: phelper-3} to observe that the total credits distributed by each batch $X$ of type $k$ scheduled by $\hOPT$ is at most $2^{k + 2}$, while the cost incurred by $\hOPT$ for using batch $X$ is $2^k$. Therefore, the total amount of credit awarded for all batches used by $\hOPT$ will not exceed $4 \cdot \cost(\hOPT)$.
    This finishes the proof of Lemma \ref{lem: pvalid-assgn-lb}.
\end{proof}

The constant of $1/4$ given by Lemma \ref{lem: pvalid-assgn-lb}
is tight even for $p=1$.  See Appendix \ref{sec: app-tight-lb}.

\subsection{The core algorithm and its analysis}\label{sec: palgs}

\begin{figure}
    \centering
    \includegraphics[width = 13cm]{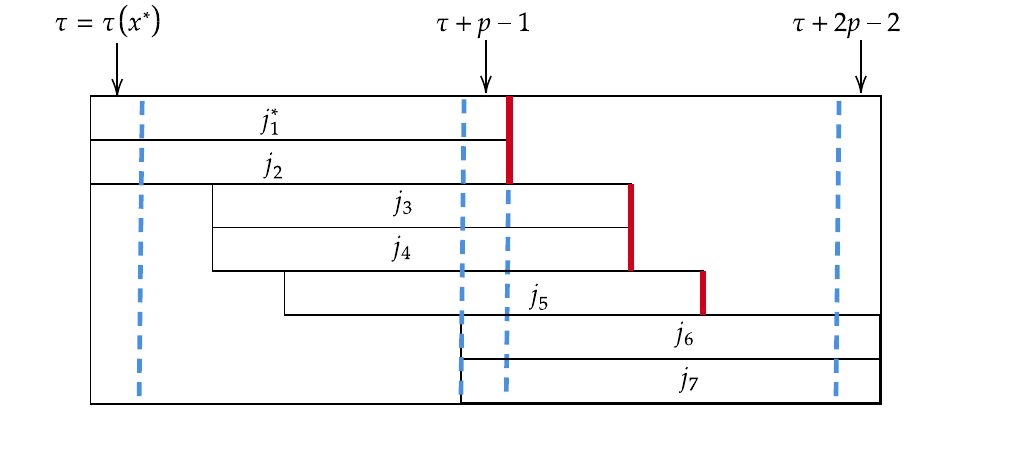}
    \caption{\small{An example of a batch $X^*$ as finalized by Algorithm \ref{alg: pmain}. Based on its type 
    (as computed by Algorithm \ref{alg: ptype}, computation not shown in this figure),
    the capacity of this batch is 7.
    $X^*$ is created at time $\tau(X^*) = \tau$ when  job $j^* = j^*_1$ has   
    reached $\tau = d_{j^*} - p + 1$.
    Later, $j_2$ which has the same deadline as $j^*_1$ is assigned to $X^*$. 
    As $j_3, j_4$, and $j_5$ reach their starting deadlines, they are assigned to $X^*$.
    At time $\tau + p - 1$, the midpoint of $X^*$, we still have capacity to spare and $j_6$ and $j_7$
    are assigned to $X^*$ by the EDF rule. For the depicted jobs, we mark 
    their deadlines with red.
    } }
    \label{fig: batch}
\end{figure}

\begin{figure}
    \centering
    \includegraphics[width = 13cm]{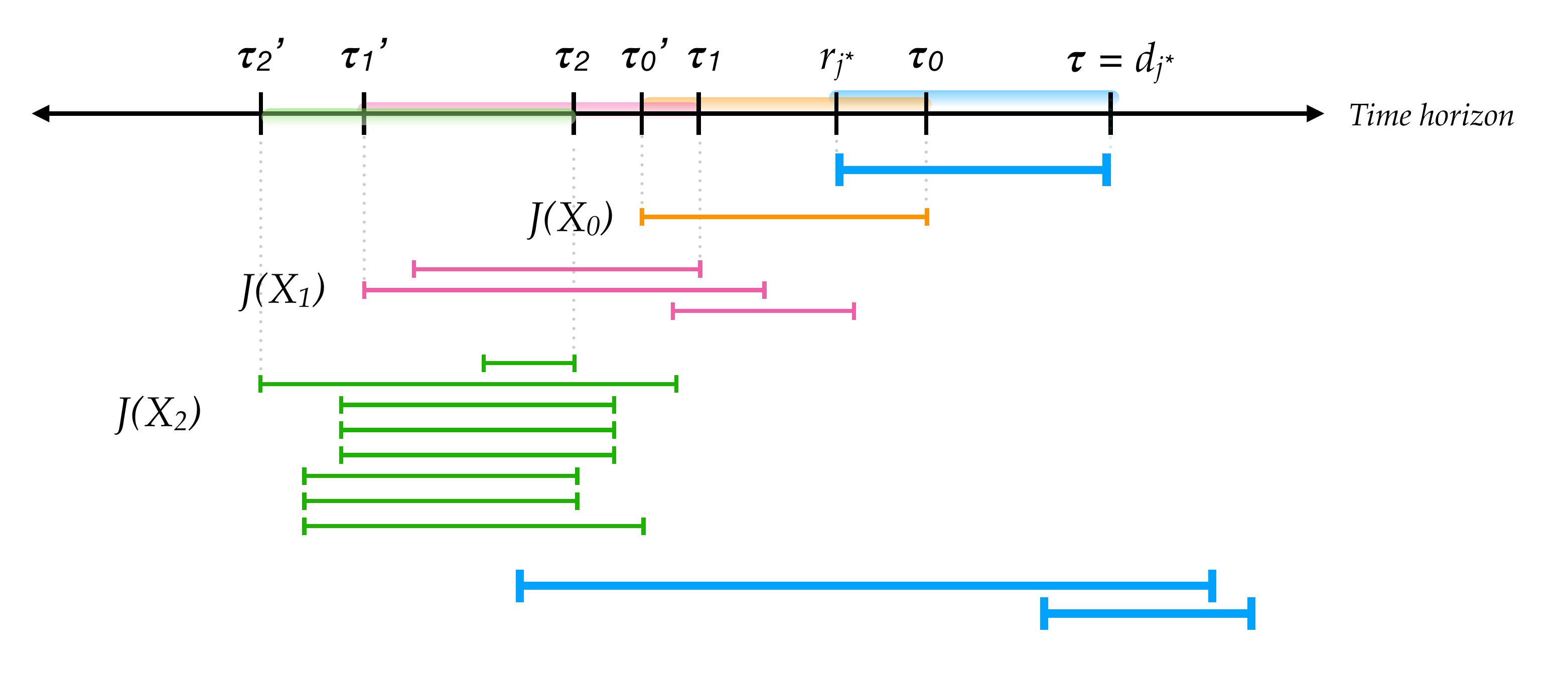}
    \caption{\small{Constructing intervals and assigning jobs to the new batch.
    In this figure, we have unit-processing-time jobs. 
    Then the start of a batch equals its midpoint equals its end, and all action takes place at time  $\tau$.
    At time $\tau = d_{j^*}$, job $j^*$ (highest blue interval giving the release time and deadline of $j^*$)
    must be executed.  
    Algorithm \ref{alg: ptype} decides which machine type to use for the next batch at $\tau$ by constructing intervals $I_0,I_1,I_2,I_3$, 
    which are indicated on the time horizon, where $I_0$ is the blue interval, $I_1$ is the union of the blue and orange intervals, $I_2$ is the union of the blue, orange, and pink intervals, and $I_3$ is the union of all colored intervals. These intervals are defined by taking the previous interval and extending it to the earliest release time of a job in the corresponding batch.
    Here Algorithm \ref{alg: ptype} decides to use a batch of type 3.
    Capacities of batches in this instance are $B_0=1$, $B_1=3$, $B_2 = 8$, $B_3 = 20$. Jobs with blue (release time, deadline) intervals are executed in the new batch, which can have fewer than $B_3$ jobs if these are all
    the jobs in the waiting list $W$.} }
    \label{fig:intervals}
\end{figure}

\begin{figure}
    \centering
    \includegraphics[width = 13cm, angle = 0]{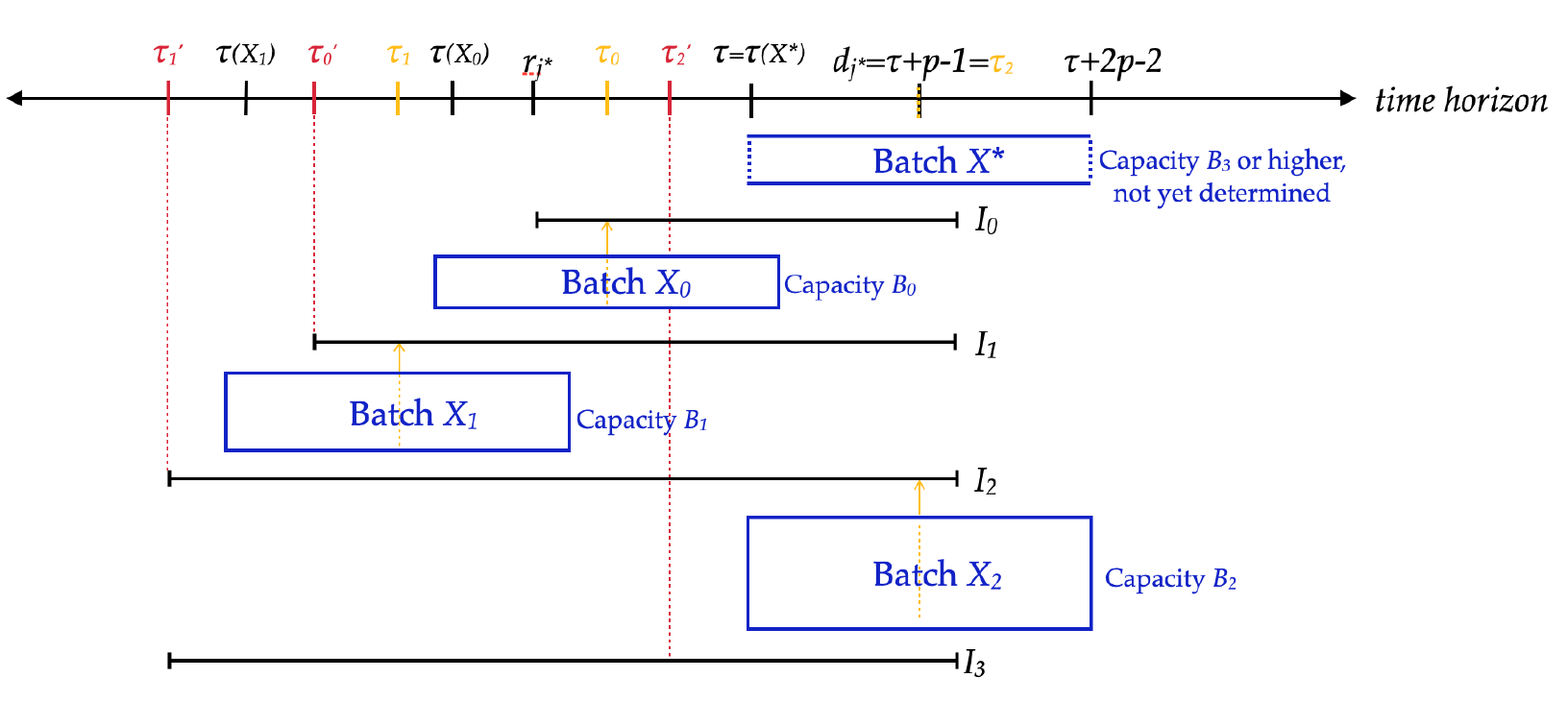}
    \caption{\small{A partial run of Algorithm \ref{alg: ptype}. We are at time $\tau$ and
    are choosing the type for a new batch $X^*$ that is to run from $\tau$ to $\tau + 2p-2$,
    with midpoint $\tau + p - 1$.
    This batch is created by job $j^*$ with deadline $d_{j^*} = \tau + p -1$.
    The interval $I_0 = [r_{j^*}, d_{j^*}]$.
    Here, there exists a batch $X_0$ with its
    midpoint $\tau(X_0) + p - 1$ inside $I_0$ (midpoints of batches are marked with yellow dotted lines). Thus $I_1$ is created with left endpoint
    the earliest release time of a job in $J(X_0)$ (the earliest release time of a job in $J(X_\ell)$ is $\tau'_\ell$, for $\ell \in \{0,1,2\}$). Later, $I_2$ and $I_3$ are created as depicted, and Algorithm \ref{alg: ptype} is not finished yet.
    } }
    \label{fig:alg3}
\end{figure}

We use Algorithm \ref{alg: pmain} to prove Lemma \ref{lem: main}.  At any time during the algorithm's execution, there is at most one batch $X^*$ (and hence at most one machine) that is open to processing more jobs if it has excess capacity.
The algorithm scans the time horizon until we reach a timeslot
$\tau$ where  either some job $j^*$ is not assigned and has reached $\tau = d_{j^*} - p + 1$,
or the batch $X^*$ reaches the midpoint of its active interval.
If there is a job $j^*$ as above  and $X^*$ is not full (in other words,
$|J(X^*)| < B_{t(X^*)}$, where $t(X)$ is the type of batch $X$),
then assign $j^*$ to batch $X^*$.
If there is a job $j^*$ as above  and $X^*$ is full,
Algorithm \ref{alg: pmain} must decide which machine type to open and use for $j^*$,
and this decision is made by the subroutine in Algorithm \ref{alg: ptype}.
Examples of partial runs for Algorithm \ref{alg: pmain} and Algorithm \ref{alg: ptype}
appear in Figures \ref{fig:intervals} and  \ref{fig:alg3}. 
If the batch $X^*$ reaches the midpoint of its active interval,
then we fill it up to capacity (or we fill it up with all the waiting jobs)
using the EDF rule. An illustration of this process is in Figure \ref{fig: batch}.
The variable $W$ is the set of unassigned jobs  (jobs waiting to be scheduled). 

Recall that in the proof of the lower bound, it was helpful to associate a type $t_I$ and a set of jobs $J_I$ to interval $I$. 
In the algorithm analysis, it will be helpful to associate a type $t(X)$, two sets of jobs $\widetilde{S}(X)$ and $J(X)$, and an interval $\widetilde{I}(X)$ to batch $X$.  
It is not surprising to see $t(X)$ as the associated type, but what might be surprising is to see two sets of associated jobs.
While $J(X)$ are the jobs in batch $X$, 
$\widetilde{S}(X)$ are the jobs we charge to $X$ for our analysis, 
where jobs may be paying for a batch other than the one they are a part of.

\begin{algorithm}[t!]
\caption{The core algorithm}
\label{alg: pmain}
\begin{algorithmic}
 \State Let $W \leftarrow \emptyset$,  $\tau \leftarrow 0$, $\mathcal{X} \leftarrow \emptyset$, and $X^* \leftarrow$ Undefined.
\While{$\tau \neq \textsf{NULL}$} 
  \State $\rhd$ Let $J'$ be the set of jobs with release time $\tau$.
  \State $\rhd$ Update $W \leftarrow W \cup J'$.
  \While{$\exists$ $j^* \in W$ with $d_{j^*} = \tau + p - 1$}
  {{\emph{ // If many such $j^*$, choose one arbitrarily}}}
     \If{ $X^*$ is defined and $|J(X^*)| < B_{t(X^*)}$}
         \State $\rhd$ $J(X^*) \leftarrow J(X^*) \cup \{j^*\}$, $W \leftarrow W \setminus \{j^*\}$.
     \Else
        \If{$X^*$ is defined} {{\emph{ // Here, the current batch $X^*$ is full}}}
            \State $\rhd$ 
            Update $\mathcal{X} \leftarrow \mathcal{X} \cup \{X^*\}$.
        This adds $\{(\widetilde{I}(X^*),t(X^*), \widetilde{S}(X^*))\}$ to $\{(\widetilde{I}(X),t(X),\widetilde{S}(X))\}_{X\in \mathcal{X}}$.
        \EndIf
          \State $\rhd$ Run Algorithm \ref{alg: ptype} with inputs $\tau$, 
          $W$, $j^*$, and $\mathcal{X}$, which
         modifies $W$ and  
          outputs new batch $X^*$ 
          equipped with $t(X^*)$, $J(X^*),\tau(X^*), \widetilde{S}(X^*$), $\widetilde{I}(X^*)$.
     \EndIf
  \EndWhile
    \If{ $X^*$ is defined and $\tau = \tau(X^*) + p - 1$}
        \While{$|J(X^*)| < B_{t(X^*)}$ and $|W| > 0$} 
            \State $\rhd$  Find the job $j  \in W$ with earliest deadline.
            \State $\rhd$ Update  $J(X^*) \leftarrow J(X^*) \cup \{j\}$, $W \leftarrow W \setminus \{j\}$.
        \EndWhile
        \State $\rhd$ 
        Update $\mathcal{X} \leftarrow \mathcal{X} \cup \{X^*\}$.
        This adds $\{(\widetilde{I}(X^*),t(X^*), \widetilde{S}(X^*))\}$ to $\{(\widetilde{I}(X),t(X),\widetilde{S}(X))\}_{X\in \mathcal{X}}$.
        \State $\rhd$ $X^* \leftarrow$ \textsf{Undefined}.
    \EndIf
    
  \State $\rhd$ Increment $\tau \leftarrow \tau + 1$, or let $\tau= \textsf{NULL}$ if no more arrivals and past the last deadline.
\EndWhile
      \State $\rhd$ \textbf{Return} $\mathcal{X}$.
  \end{algorithmic}
\end{algorithm}

Intuitively speaking, we employ a type of ``pay-it-forward" philosophy.
We charge the cost of one of the batches that the algorithm produces
to the cost that $\hOPT$ expends to execute
one job in that batch (which we will refer to as the critical job) and 
the non-critical jobs in another batch.
The \emph{critical job} in batch $X^*$ is the job 
that was first introduced in $J(X^*)$ when batch $X^*$ is created
in Algorithm \ref{alg: ptype}.
To determine $\widetilde{S}(X^*)$, throughout the algorithm
we will keep track of an interval associated with batch $X^*$,
which we denote by $\widetilde{I}(X^*)$.
Note this is a slight abuse of notation from the lower bound subsection, 
where we let $I(j)$ denote the interval containing job $j$.
Additionally, tracking $\widetilde{S}(X^*)$ and $\widetilde{I}(X^*)$
is not necessary for the algorithm
(hence they are given the additional tilde in the notation),
but for clarity in the analysis, we construct them in Algorithm
\ref{alg: ptype} and keep them in Algorithm \ref{alg: pmain}. 

\begin{algorithm}[t!]
\caption{Deciding the next batch}
\label{alg: ptype}
\begin{algorithmic}
 \State \textbf{Input:} time $\tau$, 
 waiting jobs $W$, critical job $j^*$ with $d_{j^*}=\tau + p - 1$, and batches $\mathcal{X}$, where every $X \in \mathcal{X}$ is equipped with $t(X)$, $J(X)$, $\tau(X)$, $\widetilde{S}(X$), $\widetilde{I}(X)$.
 \State $\rhd$ Let $k \leftarrow 0$. 
 \State $\rhd$ Let $I_0 \leftarrow [r_{j^*},\tau + p -1]$. 
                 {{\emph{ // Note that $d_{j^*} = \tau + p - 1$}}}
  \While{$I_k$ contains a timeslot $\tau_k:=\tau(X_k) + p -1$ with $t(X_k) = k$ for some $X_k \in \mathcal{X}$ }
  \State \hspace{-7mm}{{\emph{ // Only the latest such $\tau_k$ is stored and this is how $X_k$ is selected}}}
                \State $\rhd$ Let $j_k$ be a job in $J(X_k)$
                 with the earliest release time.  
                 {{\emph{ // Tie-break arbitrarily}}}
                \State $\rhd$ Let $\tau'_k$ be the release time of $j_k$.
                \State $\rhd$ Set $I_{k+1} \leftarrow I_k \cup [\tau'_k,\tau_k]$. 
                \State $\rhd$ Increment $k \leftarrow k+1$.
    \EndWhile
\State $\rhd$  $S^* \leftarrow \{j^*\}$.
    \If{$k > 0$}
    \State Add all the non-critical jobs of $X_{k-1}$ to $S^*$.
   \EndIf
   \State $ \rhd $ Create batch $X^*$ with type $t(X^*) \leftarrow k$, start time $\tau(X^*) \leftarrow \tau$, and
 jobs $J(X^*) \leftarrow \{j^*\}$.
   \State $ \rhd $ $W \leftarrow W \setminus \{j^*\}$; Define $\widetilde{I}(X^*) \leftarrow I_k$ and $\widetilde{S}(X^*) \leftarrow S^*$.
   \State {\textbf{Return} $(X^*,W)$.} // $W$ was modified
\end{algorithmic}
\end{algorithm}

In order to use our lower bound from the previous subsection, 
we will use the batches to construct a valid interval assignment (as in Definition \ref{def: pcharacterization}). 
Visualizations of the intervals Algorithm \ref{alg: ptype} constructs
are  provided in Figures \ref{fig:alg3} and \ref{fig:intervals}.
Note that Algorithm \ref{alg: pmain} 
sets $t_{\widetilde{I}(X)} = t(X)$ when it defines
$\{(\widetilde{I}(X),t(X),\widetilde{S}(X))\}_{X \in \mathcal{X}}$
as an interval assignment.
We prove that the conditions for Definition \ref{def: pcharacterization} hold in 
Lemmas \ref{lem: pcond-1-2} and \ref{lem: pcond-4} and 
Proposition \ref{prop: pcond-3}.

\begin{lemma}\label{lem: pcond-1-2}
Algorithm \ref{alg: pmain} produces 
$\{(\widetilde{I}(X),t(X),\widetilde{S}(X))\}_{X \in \mathcal{X}}$
that satisfy Conditions (1) and (2) of being a valid interval assignment (Definition \ref{def: pcharacterization}).
\end{lemma}
\begin{proof}
Let $X \in \mathcal{X}$ and let $\tau = \tau(X)$.
If $t(X) = 0$, then indeed there is only the critical job in $\widetilde{S}(X)$, so $|\widetilde{S}(X)| = 1$. 
Further, the critical job $j^*$ completely determines the interval, 
with $\widetilde{I}(X) = [r_{j^*}, d_{j^*}]$.
Thus, Conditions (1) and (2) are satisfied when $t(X)=0$.

Fix $t(X) =k>0$. 
The midpoint of batch $X$ (timeslot $\tau(X) + p - 1 = \tau + p -1$) is the right endpoint of $\widetilde{I}(X)$.
Consider the sequence of nested intervals that the algorithm built
up to construct $\widetilde{I}(X)$, i.e., 
$I_0, I_1,\ldots, I_k$, where now $\widetilde{I}(X) = I_k$.
These are indeed nested intervals, since  for all $h = 0, \ldots, k-1$,
$I_h$ contains $\tau_h$ and $I_{h+1}$ is defined by
$I_{h+1} \leftarrow I_h \cup [\tau'_h,\tau_h]$. 
Associated with these intervals are batches $X_0,X_1,\ldots,X_{k-1} \in \mathcal{X}$
that forced interval $\widetilde{I}(X)$ to be created  (see Algorithm \ref{alg: ptype}).
As in this algorithm, $\tau'_\ell$ denotes the release time
of an earliest arriving job in $J(X_\ell)$.
Also note that $t(X_\ell) = \ell$.
Since $\tau \in \widetilde{I}(X)$, 
we have that for all $\hat{\tau}  \in \widetilde{I}(X)$, either $\hat{\tau} \in I_0$ 
or there exists some $0 \leq \ell < k$ such that $\hat{\tau}  \in [\tau'_\ell,\tau_\ell]$.

\medskip

We prove by induction on $\ell <k$ that $X_\ell$ is a full batch
(that is, $|J(X_\ell)| = B_{t(X_\ell)}$), and
all jobs in $J(X_\ell)$ have deadline no later than $\tau + p - 1$.

\noindent \textbf{Base case:} Here, $\ell=0$. 
As in the algorithms, $j^*$ is the critical job of $X$.
Note that $\tau(X_0) + p -1 = \tau_0 \in I_0$,
the left endpoint of $I_0$ is $r_{j^*}$ 
and the right endpoint of $I_0$ is $\tau + p  - 1 = d_{j^*}$.
The existence of $X_0$ implies that $\tau(X_0) \leq \tau$,
and furthermore  the construction of $X_0$ is completed
at or before time $\tau$. Also, $\tau_0 \in I_0$ and therefore $r_{j^*} \leq \tau_0$.
We have $j^* \not \in J(X_0)$.
If $r_{j^*} \leq \tau(X_0) = \tau_0 - p + 1$, then 
$j^*$ was eligible to be put in $J(X_0)$ and scheduled to be executed
in timeslots $\tau_0 - p + 1, \ldots, \tau_0$ ($j^*$ would be in $W$ at time $\tau(X_0)$).
If $r_{j^*} > \tau(X_0) = \tau_0 - p + 1$, then 
$j^*$ was eligible to be put in $J(X_0)$ and could be scheduled to be executed
in timeslots $r_{j^*}, \ldots, r_{j^*} + p - 1$ ($j^*$ would be in $W$ at time $r_{j^*}$)
as $r_{j^*} + p - 1 \leq \tau_0 + p - 1 = \tau(X_0) + 2p-2$.
The fact that $j^*$ was not put by Algorithm \ref{alg: pmain} in $J(X_0)$
at either its starting deadline $d_{j^*} - p + 1$ or at the midpoint of $X_0$
(whichever comes first), implies that $|J(X_0)| = B_0$
and moreover, using the way the algorithm works and the EDF rule, 
all the jobs in $J(X_0)$ have deadline no later than $d_{j^*} = \tau + p -1$.

\noindent \textbf{Induction step:} 
Fix some integer $ 1 \leq \ell < k$.
Assume that for all $0 \leq \ell' < \ell$,
$X_{\ell'}$ is filled to its capacity, and
all the jobs in batch $X_{\ell'}$ have deadline no later than $\tau + p - 1$.
We will show that $X_{\ell}$ is filled to its capacity, and
all the jobs in batch $X_{\ell}$ have deadline no later than $\tau + p - 1$.
Since $X_\ell$ exists,
either $\tau_\ell \in I_0$ (in which case the reasoning from the base case applies),
or there exists some $\lambda < \ell$ 
with  $\tau_\ell \in [\tau'_\lambda,\tau_\lambda]$. 
The latter case is the interesting one, so we fix the smallest such $\lambda$.
Note that this implies that $\tau_\ell < \tau_\lambda$, as argued next.
Indeed, assume for a contradiction that $\tau_\ell = \tau_\lambda$.  
We have $\tau_\lambda \in I_\lambda =  \cup_{i=0}^{\lambda - 1} [\tau'_i, \tau_i] \cup I_0$.
If $\tau_\lambda \in I_0$, then $\tau_\ell \in I_0$, a case solved earlier, and 
if $\tau_\lambda \in [\tau'_i, \tau_i]$ for $i< \lambda$, then we contradict the choice of $\lambda$.
The fact that $\tau_\ell < \tau_\lambda$ implies that $\tau(X_\ell) < \tau(X_\lambda)$,
which implies that
batch $X_\ell$ was created and completely constructed before batch $X_\lambda$ is created.

As in Algorithm \ref{alg: ptype}, job $j_\lambda$ is a job arriving at time $\tau'_\lambda$ that is
scheduled in batch $X_\lambda$.
We have $j_\lambda \not \in J(X_\ell)$.
If $r_{j_\lambda} \leq \tau(X_\ell) = \tau_\ell - p + 1$, then 
$j_\lambda$ was eligible to be put in $J(X_\ell)$ and scheduled to be executed
in timeslots $\tau_\ell - p + 1, \ldots, \tau_\ell$ ($j_\lambda$ would be in $W$ at time $\tau(X_\ell)$).
If $r_{j_\lambda} > \tau(X_\ell) = \tau_\ell - p + 1$, then 
$j_\lambda$ was eligible to be put in $J(X_\ell)$ and
could be scheduled to be executed
in timeslots $r_{j_\lambda}, \ldots, r_{j_\lambda} + p - 1$ ($j_\lambda$ would be in $W$ at time $r_{j_\lambda}$)
as $r_{j_\lambda} + p - 1 \leq \tau_\ell + p - 1 = \tau(X_\ell) + 2p-2$.
The fact that $j_\lambda$ was not put by Algorithm \ref{alg: pmain} in $J(X_\ell)$ implies that $|J(X_\ell)| = B_\ell$
and moreover, using the way the algorithm works and the EDF rule, 
all the jobs in $J(X_\ell)$ have deadline no later than $d_{j_\lambda}$.
The induction hypothesis gives us that $d_{j_\lambda} \leq \tau + p - 1$,
and therefore 
all the jobs in $J(X_\ell)$ have deadline no later than $\tau + p -1$.
This completes the inductive step.

\smallskip 

Finally, let us look at the moment that the algorithm has found an interval 
$I_k$ that does not contain a time $\tau_k$;
we can assume that $k > 0$ as we already considered the case when $k=0$.
Recall that this will be the interval associated with the new batch,
i.e., the algorithm sets $\widetilde{I}(X) = I_k$.
Note that $[\tau'_{k-1},\tau + p - 1] \subseteq I_k$ and recall that 
 $\widetilde{S}(X)$ is the critical job of batch $X$
plus all the non-critical jobs of batch $X_{k-1}$.
Since batch $X_{k-1}$ is full, 
$|\widetilde{S}(X)| = B_{k-1}$ and so Condition (1) is satisfied.
Moreover, every job in $J(X_{k-1})$ has release time 
at or after $\tau'_{k-1}$ and deadline at most $\tau + p - 1$, as proven above. 
The critical job of batch $X$
also has release time and deadline inside $I_k$ (since $I_k$ contains $I_0$),
and we conclude that Condition (2) is satisfied since $\widetilde{I}(X) = I_k$.
\end{proof}

\begin{proposition}
    \label{prop: pcond-3}
Algorithm \ref{alg: pmain} produces batches $\mathcal{X}$ 
equipped with 
$\{(\widetilde{I}(X),t(X),\widetilde{S}(X))\}_{X \in \mathcal{X}}$ 
which satisfy Condition (3) of being a valid interval assignment (Definition \ref{def: pcharacterization}).
\end{proposition}
\begin{proof}
Note that the right endpoint of $\widetilde{I}(X)$ is $\tau(X) + p - 1$.
Suppose that in Algorithm \ref{alg: ptype}, $t(X^*)$ is set to $k$. 
If $X_k$ is as in Algorithm \ref{alg: ptype}, then 
$\tau(X_k) \leq \tau =  \tau(X^*)$ ($X^*$ is created after $X_k$)
and therefore $\tau_k$ is at most the right endpoint of $I_k$.
So if $\tau_k \not \in I_k$, it is because $\tau_k$ is before
the left endpoint of $I_k$. Note that $\widetilde{I}(X_k)$ 
is an interval with right endpoint $\tau_k$ and therefore
it is disjoint from $I_k = \widetilde{I}(X^*)$.

\end{proof}

\begin{lemma}\label{lem: pcond-4}
Algorithm \ref{alg: pmain} produces batches $\mathcal{X}$  equipped with
$\{(\widetilde{I}(X),t(X),\widetilde{S}(X))\}_{X \in \mathcal{X}}$ 
which satisfy Condition (4) of being a valid interval assignment 
(Definition \ref{def: pcharacterization}).
\end{lemma}
\begin{proof}
Looking at the algorithm, 
a job $j$ critical for batch $X$ is put in $\widetilde{S}(X)$ and no other $\widetilde{S}(X')$  for $X' \neq X$.
Any other job $j'$ is a non-critical job in $J(X)$, for some batch $X$.
Let $k = t(X)$, the type of $X$. If $j'$ were
in $\widetilde{S}(X_1)$ and $\widetilde{S}(X_2)$, for batches $X_1 \neq X_2$, then
both $X_1$ and $X_2$ are of type $k+1$. 
But then, by Condition (3), these corresponding intervals 
$\widetilde{I}(X_1)$ and $\widetilde{I}(X_2)$ must be disjoint.
These intervals cannot be disjoint, since by Condition (2)
they are both required to contain $[r_{j'}, d_{j'}]$.
\end{proof}

\begin{proof}[\emph{Lemma \ref{lem: main}}]
Combining Lemmas \ref{lem: pcond-1-2} and \ref{lem: pcond-4} with Proposition \ref{prop: pcond-3}, 
we see that Algorithm \ref{alg: pmain} schedules batches $\mathcal{X}$ 
 with a valid interval assignment 
$\{(\widetilde{I}(X),t(X),\widetilde{S}(X))\}_{X \in \mathcal{X}}$.
Recall that the costs of machines are all powers of 2 and all machines have to be used for exactly $2p-1$
timeslots. 

The cost of the batches in $\mathcal{X}$ is  $$\sum_{ X \in \mathcal{X}} 2^{t(X)}.$$
Applying Lemma \ref{lem: pvalid-assgn-lb} to valid interval assignment
$\{(\widetilde{I}(X),t(X),\widetilde{S}(X))\}_{X \in \mathcal{X}}$,
we see that for $\cL = \{\widetilde{I}(X)\}_{X \in \mathcal{X}}$:
$$\cost(\hOPT) \geq \frac14 \cdot  \sum_{I \in \cL} 2^{t_I}  = \frac{1}{4} \cdot  \sum_{X \in \mathcal{X}} 2^{t(X)} .$$

Therefore, the cost of $\mathcal{X}$ is at most $4 \cdot \cost(\hOPT)$.

The running time of this algorithm is $O(n \log n)$. 

Offline, as in Algorithm \ref{algo:greedy},
we sort $J$ so that $J'$ can be found in time $O(|J'|)$.
Online,  arrivals are already revealed chronologically.
We  use a min-heap with keys $d_j -p + 1$,
for elements $j \in W$, to store $W$.
We use the sorted
$J$ to find the next timeslot where $J' \neq \emptyset$ and compare
it with the smallest key in $W$  and with $\tau(X^*) + p -1$.
Recall that
$K \leq 2 + \log_2 n$ is an upper bound on the number of types of machines 
by Proposition \ref{prop:pUB-K}.
We also keep for each type  $\ell \in \{0, \ldots, K-1\}$
the time $\tau(X)$ and the earliest
release time of a job in $J(X)$, where $X$ is the latest batch of type $\ell$.
Using these,
each critical job requires at most $K+1$ iterations through the while loop
of Algorithm \ref{alg: ptype}
to choose the type of the batch which will contain this critical job.
In time $O(\log n)$ per job,
we can update the data structures.
\end{proof}

\section{Lower Bounds on the Competitive Ratio} \label{sec: adversary-4}

In this section we prove Theorems \ref{thm: lb_nested_deterministic} and \ref{thm: lb_nested_randomized}.
We use results and techniques from \cite{SLLA25}, with non-trivial adaptations for our problem. 
After we obtained our reduction, we noticed similarities with
the constructions of  \cite{LT24}
 and the v1 version of \cite{albers2025onlinebusytimescheduling}
 (both claiming a lower bound of $4$ for  deterministic 
 online algorithms for Span Minimization, which is Busy Time
 with flexible jobs of arbitrary processing time and one machine type of
 unbounded capacity).
 
Recall that these results are stated for $p=1$. 
First, we follow \cite{SLLA25} by defining an online \emph{Button  problem}.

In this problem, we are given an ordered list $b$ of $m+1$ buttons where each button $j$ is associated with a price $b_j \in \QQ_{\geq 1}$.
The prices are monotone: $1 = b_0 \leq b_1 \leq b_2 \leq \cdots \leq b_m$.  Some buttons are designated as target buttons, which form a
suffix of the button list, i.e., there exists some $j^* \leq m$ such that buttons $j^*$ to $m$ are all targets, and
none of the  buttons before $j^*$ are targets. The list $b$ is known from the start.
However, we do not know in advance the first target button
(i.e., $j^*$). We can learn whether a button $j$ is a target or not only by pressing the button at the price
of $b_j$. The online algorithm repeatedly presses buttons until it presses any target button, and the natural
objective is to minimize the total price.

We say that deterministic online algorithm $\cA$ is $\gamma$-competitive if for all lists $b$ and all 
$j^* \in \{0, 1, \ldots, m\}$, $\cost(\cA,b,j^*)  \leq \gamma \cdot b_{j^*}$,
where $\cost(\cA,b,j^*)$ is the total price that the algorithm $\cA$ incurs on input $b$
until it presses a target button when the first target button is $j^*$. 
Note that algorithm $\cA$ can be identified with a sequence  $i_1(\cA,b), i_2(\cA,b), \ldots, i_{q(\cA,b)}(\cA,b)$
which are the buttons pressed by the algorithm  when the target is $m$, 
where we assume without loss of generality that $i_j(\cA,b) < i_{j+1}(\cA,b)$
and $i_{q(\cA,b)}(\cA,b) =  m$. 
The algorithm, of course, stops after hitting the first target, i.e. the first $l = l(\cA,b,j^*)$
for which $i_l(\cA,b) \geq j^*$, and its total price is $\cost(\cA,b,j^*) = \sum_{h=1}^l b_{i_h(\cA,b)}$.

Combining Lemma 5.9 and Theorem 5.11 of \cite{SLLA25}   into a single theorem, we have:
\begin{theorem}\label{thm: lb_button_4} [from Section 5.3 of \cite{SLLA25}]
    For any $\epsilon > 0$ there is an $m = m(\epsilon)$ and a list $b$ of button prices $ 1 = b_0 < b_1 \cdots < b_m$ such that
    for any deterministic online
    algorithm $\cA$ there exists a $j^* = j^*(\epsilon,\cA)$ such that
    $\cost(\cA,b,j^*) > (4 - \epsilon) b_{j^*}$.
    Rephrased:
    For any $\epsilon > 0$ there is an $m = m(\epsilon)$ 
    and a list $b$ of button prices $ 1 = b_0 < b_1 < \cdots < b_m$
    such that
    for any increasing sequence $S$ of button indices $i_1, i_2, \ldots, i_k = m$,
    there exists a  $j^* = j^*(\epsilon,S)$ such that
    $ \sum_{h=1}^v b_{i_h} > (4 - \epsilon) b_{j^*}$, where
    $v = \min \{u : {i_u} \geq j^*\}$.

\end{theorem}
This theorem implies that no deterministic online algorithm for the Button problem
has a competitive ratio smaller than 4.
The statement of the theorem is stronger than this, however, as the list $b$ does not depend on the algorithm $\cA$; we will use this fact.
As an aside, the list $b$ is given by $b_i = 1 + i \cdot \delta$, for a small $\delta$ that depends on $\epsilon$.
See \cite{SLLA25} for details. Earlier work by \cite{zhang2011ski} proves a very
similar statement for Multi Option Ski Rental,  and they explicitly notice that the same list of options works
against all deterministic online algorithms. We prefer to use \cite{SLLA25}  instead of \cite{zhang2011ski} for simplicity, and 
for its discussion on randomized algorithms.

\begin{proof}[\emph{Theorem \ref{thm: lb_nested_deterministic}}]
   Let $\epsilon > 0$.  We pick a large integer $N = N(\eps)$ whose exact value will be fixed later. 
    Let $m = m(\eps/4)$, and let list $b =b (\eps/4)$ be from Theorem \ref{thm: lb_button_4}, creating
    an instance of the Button problem.
    Let $C= C(\eps) = \lceil \frac{16}{\epsilon}\cdot b_m \rceil$. 
        We have $m+2$ machine types, indexed from $0$ to $m+1$. For $i \in \{0,\ldots,m\}$,
         machine type $i$ has cost $c_i = b_i$ and capacity $B_i = C^i$.
        Machines of type $m+1$  have cost $c_{m+1} = c_m \cdot (1 + \eps/4)$ and capacity $B_{m+1} = N C^m$.
        
   Let $\cA$ be an arbitrary deterministic online algorithm for 
Unit-processing-time Busy Time Scheduling on Heterogeneous Machines.
The machine types do not depend on $\cA$, but the (very many) 
jobs released by the all-powerful adversary will depend on $\cA$.

The adversary will make $N$ {\em mega-releases}, each consisting of $C^m$ jobs arriving at the same time and
with different deadlines, with fixed gaps between the deadlines of jobs from the same mega-release.
We construct disjoint intervals, which we call {\em phases}, that are helpful in the proof. Phase 0 is the set of negative timeslots.

For $q =  1, 2, \ldots, N$, the adversary executes iteration $q$ of the following procedure.
At the first timeslot following phase $q-1$, the adversary initiates mega-release $q$.
For $i \in \{1,\ldots,C^m\}$, the $i^{th}$ job of mega-release $q$ has deadline
$  i (1 + C^m)^{N-q} + \sum_{l = 1}^{q-1}  C^{j_l^*} (1+C^m)^{N-l} $, 
so the gap between the deadlines of the $i^{th}$ and $(i+1)^{th}$ jobs in mega-release $q$ is $  (1 + C^m)^{N-q} $.
In iteration $q$, the adversary also chooses $j_q^*$ as explained later below. 
 Phase $q$ consists of the times from the arrival of mega-release $q$  to
 $\sum_{l = 1}^{q}  C^{j_l^*} (1+C^m)^{N-l}$.

In iteration $q$, the adversary
constructs from $\cA$ a deterministic online
algorithm $\cA_q$ for the Button problem with input list $b$.
See Figure \ref{fig: lb4} for an illustration.
Even if $\cA$ scheduled in phases $0, 1, \ldots, q-1$
all jobs of mega-releases $1, 2, \ldots, q-1$, it still needs to schedule the jobs of mega-release $q$.
Specifically, $\cA_q$ is constructed based on what $\cA$ does in the time between when mega-release $q$ arrives to the moment it
finishes scheduling all jobs of mega-release $q$ (assuming no other jobs are released):
\begin{itemize}
    \item  $\cA_q$ always starts with the button corresponding to the machine that $\cA$ would use for
scheduling the first job of mega-release $q$, except 
    if $\cA$ uses type $m+1$, then $\cA_q$ presses button $m$.
    \item After the start, if algorithm $\cA$ uses a machine of type $i$, where type $i$ is higher than all types used previously,
then $\cA_q$ presses button $i$, except
    if $\cA$ uses type $m+1$, then $\cA_q$ presses button $m$.
\item 
If $\cA$ uses at least $C$ machines,
then $\cA_q$ stops following $\cA$ and presses button $m$. 
\end{itemize}

Note that, if $\cA$ uses fewer than $C$ machines, it will have to use a machine of type $m$ or $m+1$
to schedule all the $C^m$ jobs of the mega-release $q$.  
As a result, $\cA_q$ is a valid online algorithm for the Button problem.
$\cA_q$ does not need to be the same algorithm as $\cA_{q'}$, as $\cA$ can treat mega-releases differently.
Nevertheless, $\cA_q$ is a deterministic
online algorithm for the same Button instance, and the adversary knows  how $\cA_q$ operates
after making mega-release $q$.
From Theorem \ref{thm: lb_button_4}
there exists a $j^*_q$ for which $\cost(\cA_q,b,j^*_q) > (4-\eps/4) b_{j^*_q}$.
This finishes the description of the adversary's choices in iteration $q$.

Note that no job in mega-release $q$ can be scheduled in any of phases $0, 1, \ldots, q-1$.
Also, for $i >0$, we have 
$  C^m (1 + C^m)^{N-q} +   \sum_{l = 1}^{q-1}  C^{j_l^*} (1+C^m)^{N-l}  < (1 + C^{j_{q-i}^*}) (1+C^m)^{N-(q-i)} + 
 \sum_{l = 1}^{q-i-1}  C^{j_l^*} (1+C^m)^{N-l}$ and therefore
every job in mega-release $q$ is either nested in or disjoint from every job of mega-release $q-i$.

\begin{figure}[th]
\begin{center}\leavevmode%
\scalebox{0.58}{
  \includegraphics{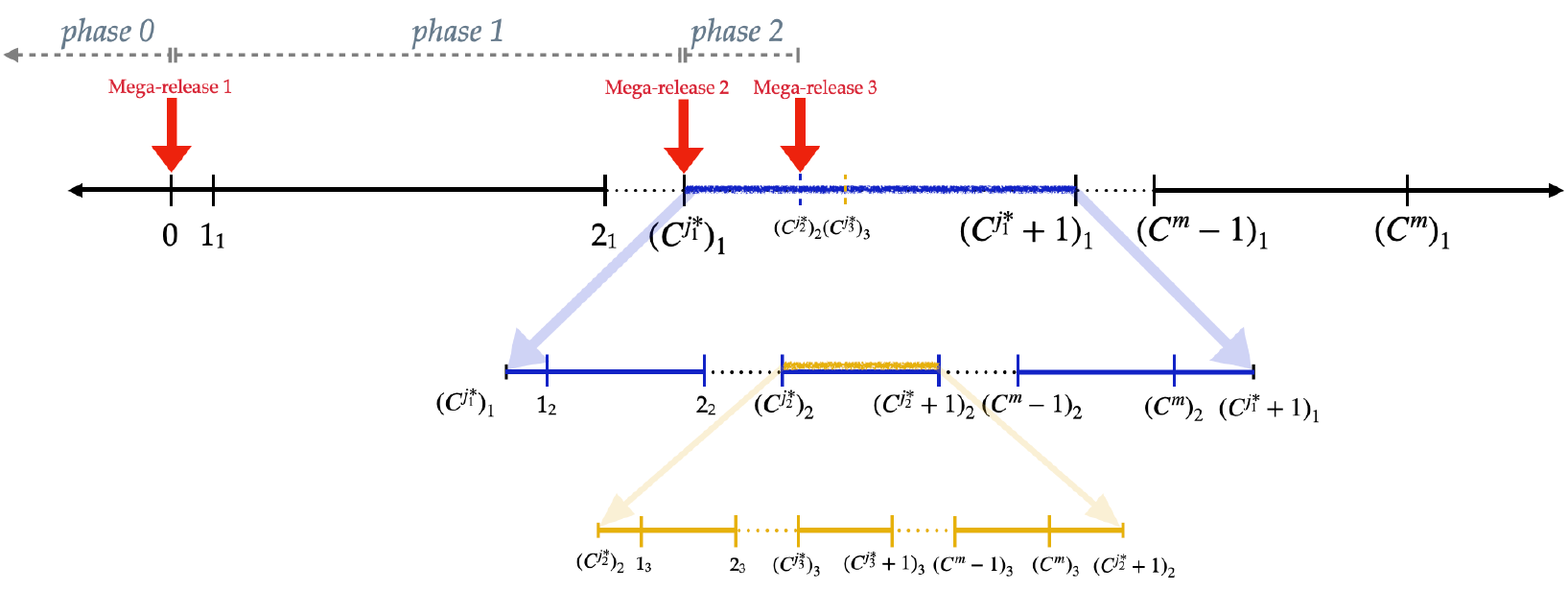}
}
\end{center}
\caption{An illustration of the construction.  We use $i_q$ to depict the deadline of the $i^{th}$ job of mega-release $q$.
All the jobs of mega-release 2 are nested between
the deadline of the $(C^{j^*_1})^{\text{th}}$ job of mega-release 1 and 
the deadline of  job $1+C^{j^*_1}$ from mega-release 1,
except if $j^*_1 = m$, when  all jobs of mega-release 2 are released
after the deadlines of all jobs of  mega-release 1. 
We have the same situation for the jobs of mega-release 3, nested 
between two consecutive deadlines of the jobs of mega-release 2 (with the same exception).
}
\label{fig: lb4}
\end{figure}

We achieved that  the $N+1$ phases occupy disjoint timeslots. For $i = 1, 2, \ldots, N$
we get deterministic
online algorithms $\cA_1, \cA_2, \ldots, \cA_N$ for the Button problem with input $b$,
and integers $j^*_1, j^*_2, \ldots, j^*_N$ such that
 $\cost(\cA_i,b,j^*_i) > (4-\eps/4) \cdot b_{j^*_i}$.

Note that $\cA$ must schedule the  $(C^{j^*_i})^{th}$ 
job of mega-release $i$ during phase $i$, because of the deadline of this job. To schedule $C^{j^*_i}$ jobs, $\cA$ must either
use a machine of type at least $j^*_i$ and capacity at least 
 $C^{j^*_i}$, or use at least $C$ machines (since  
 type $h$ for $h < j^*_i$ has capacity at most $C^{j^*_i - 1}$).

Using the previous paragraph,
the cost of $\cA_i$ on Button instance  $b$ with $j^*_i$ as the first target 
is either at most the cost of $\cA$ in phase $i$
(in the case where $\cA$ uses fewer than $C$ machines), or at most $c_m$ plus the cost of $\cA$ in phase $i$
(in the case where $\cA$ uses at least $C$ machines).
$C$ machines cost at least $C$ (the smallest cost is $1$), and $c_m \leq (\eps/16)\cdot  C$.
So, in all cases, 
the cost of $\cA_i$ on Button instance  $b$ with $j^*_i$ is at most $(1 + \eps/16)$ times the cost
of $\cA$ in phase $i$. Using the last inequality of the previous paragraph and algebraic manipulation,
we obtain that the cost of $\cA$ in phase $i$ exceeds $(4 - \eps/2) \cdot b_{j^*_i}$.
Therefore:
\begin{equation}\label{eq_sigma}
   \cost(\cA)  > (4 - \eps/2) \cdot \sum_{i=1}^N  b_{j^*_i}.
\end{equation}
Now we construct another (offline) solution $S$ to this  instance of 
Unit-processing-time Busy Time Scheduling on Heterogeneous Machines:
for all $i = 1, 2, \ldots, N$,
use one machine of type $j^*_i$ at the beginning of phase $i$, and after the end of phase $N$, use
one machine of type $m+1$. One can check that this indeed schedules all the jobs of this instance.

Now we pick $N \geq \lceil \frac{8}{\eps} c_{m+1} \rceil$.
Since $b_{j^*_i} \geq b_0 = 1$, we get $N \leq \sum_{i=1}^N b_{j^*_i}$.
Using this and algebraic manipulation, we obtain that the cost of $S$ 
is at most $\sum_{i=1}^N b_{j^*_i} ( 1 + \eps/8)$.
Using Inequality (\ref{eq_sigma}) and algebraic manipulation, we obtain
\[ \cost(\cA) > (4 - \eps) \cost(S),\]
  completing the proof of Theorem \ref{thm: lb_nested_deterministic}. 
\end{proof}

We say that a randomized
online algorithm for the Button problem $\cA$ is $\gamma$-competitive if for all lists $b$ and all 
$j^* \in \{0, 1, \ldots, m\}$, $\mathbb{E}[\cost(\cA,b,j^*)]  \leq \gamma \cdot  b_{j^*}$,
where $\cost(\cA,b,j^*)$ is the total price that the algorithm $\cA$ incurs on input $b$
until it presses a target button, when the first target button is $j^*$. 
Note that a randomized
online algorithm $\cA$ can be identified with a distribution over 
deterministic
online algorithms, and thus a distribution over sequences defined earlier for
deterministic online algorithms.

Recall that $e$ is Euler's number. From the proof of Theorem 5.2 of \cite{SLLA25}, we have:
\begin{theorem}\label{thm: lb_button_e} [from Section 5.3 of \cite{SLLA25}]
    For any $\epsilon > 0$ there is an $m = m(\epsilon)$ and a list $b$ of button prices
    $ 1 = b_0 < b_1 \cdots < b_m$ such that
    for any randomized
    online algorithm $\cA$ there exists a $j^* = j^*(\epsilon,\cA)$ such that
    $\mathbb{E}[\cost(\cA,b,j^*) ]> (e - \epsilon) \cdot b_{j^*}$.
    Rephrased:
    For any $\epsilon > 0$ there is an $m = m(\epsilon)$ and a list $b$ 
    of button prices $ 1 = b_0 < b_1 < \cdots < b_m$ such that
    for any distribution $\Pi$ over increasing sequences $S$ of button indices
    $i^S_1, i^S_2, \ldots, i^S_{k(S)} = m$,
    there exists a  $j^* = j^*(\epsilon,\Pi)$ such that
    $ \mathbb{E}_{S \sim \Pi} \big[  \sum_{h=1}^{v(S)} b_{i^S_h} \big] > (e - \epsilon) b_{j^*}$, where
    $v(S) =  \min \{u : {i^S_u} \geq j^*\}$.
\end{theorem}
This theorem implies that no randomized
online algorithm for the Button problem
has a competitive ratio smaller than $e$.
The statement of the theorem is stronger than this, since the list $b$ does not depend on the algorithm $\cA$ and we will use this fact.
As an aside, the list $b$ is given by $b_i = e^{i/\Delta}$, for a large $\Delta$ that depends on $\epsilon$.
See \cite{SLLA25} for details. 

\begin{proof}[\emph{Theorem \ref{thm: lb_nested_randomized}}]
    We use exactly the same construction as in the proof of Theorem \ref{thm: lb_nested_deterministic},
    except that the vector $b$ is chosen and
    $j^*_i$ is selected from Theorem \ref{thm: lb_button_e} 
    such that  $\mathbb{E}[\cost(\cA_i)] > (e - \eps/4) \cdot b_{j^*_i}$,
    where $\cA_i$ is a randomized online algorithm for the Button problem that we describe next.
    At the moment of mega-release $i$, the randomized
    online algorithm $\cA$ can be in a number of configurations,
    each with a certain probability associated with it. The oblivious (but otherwise all-powerful)
    adversary can  compute these probabilities, and then obtain a randomized
    online algorithm $\cA'_i$ based on how
    $\cA$ handles the jobs of mega-release $i$ assuming no further job releases.
    From this $\cA'_i$, the adversary can deduce $\cA_i$ 
    as in the previous proof.
    As a result, $\cA_i$ is a randomized online algorithm 
    for the Button problem that is known to the 
    adversary based on the adversary's knowledge of $\cA$ and of mega-releases $1$ through $i$,
    and so the adversary can indeed choose $j^*_i$.
    
   The $j^*_q$'s are chosen from $\cA$'s \emph{distribution}, not its realized random choices,
    so the whole instance is fixed before $\cA$ runs.
    The rest of the proof is exactly the same 
    as in the previous proof,
    but we apply Theorem \ref{thm: lb_button_e} to see that
    the expected  cost of  $\cA$ in
    phase $i$ exceeds $(e - \eps/2 ) \cdot b_{j^*_i}$.
    The expected (total) cost is at least
    the sum of expected costs over phases, and then $e$ replaces $4$ in the 
    following inequalities as  well, with the algebraic manipulations holding for $e$ as well.
\end{proof}

\section{Conclusion and Future Directions}

We study online busy time scheduling with flexible, uniform-processing-time
jobs on heterogeneous machines. For the general case, we design a 
deterministic $8(2p-1)/p$-competitive online algorithm
and a randomized
online algorithm that is $4(2p-1)/(p \ln 2)$-competitive against an oblivious adversary. When jobs have unit processing time
and agreeable deadlines, a simpler algorithm is 2-competitive, which is tight.   We focused on jobs that have uniform processing time and 
unit resource requirement\footnote{Recall that in the literature, the resource requirement is also known as the height of a job.},
which could be relaxed in future work. 
Indeed, Ren and Tang \cite{Ren-tang} consider inflexible jobs with different heights.
Sarpatwar et al. \cite{sarpatwar2023preemptive} consider preemptively scheduling flexible jobs with arbitrary processing times and heights on homogeneous machines; in the online setting, there is a $3/2$ lower bound 
on the competitive ratio of any online algorithm
that knows a job's characteristics at release time, even assuming
agreeable deadlines.

Generalizing our setting to allow non-uniform processing times or 
non-uniform heights introduces significant complexity into the problem. 
If jobs have different processing times, then we are no longer able to only schedule jobs $p-1$ timeslots before deadlines, 
and would need to reexamine our earliest deadline first heuristic. For heterogeneous machines, if jobs have unit processing time
and agreeable deadlines but are allowed to have
a resource requirement, then we can still apply our Greedy algorithm
(Algorithm \ref{algo:greedy}) and get a competitive ratio of 2, but
this may not be a polynomial-time algorithm---for example,
if jobs have heights and a machine can only process a
set of jobs of total height at most $1$ (homogeneous machines),
then $\GSS()$ must solve a bin-packing problem.
If $\GSS()$ is an $\alpha$-approximation algorithm,
we get a competitive ratio of $2 \alpha$.
When deadlines are not agreeable and jobs have heights, 
we do not know how to extend our 4-competitive algorithm 
(the core algorithm is given by Lemma \ref{lem: main}),
even when jobs are unit-processing-time. One obstacle is that
we can no longer assume that the machine type
with the largest capacity is the most cost-efficient, 
as some of the jobs may not fit on smaller machines. 

\section*{Acknowledgements} We would like to thank an anonymous referee for suggesting that our 
core algorithm for unit-processing-time 
jobs can be adapted to jobs of uniform processing time; indeed it can be adapted.

\bibliographystyle{alpha}
\bibliography{bib-2026}

\appendix

\section{Discussion on other algorithms}\label{sec: app-algs}

We investigated several simpler algorithmic approaches for the general (non-agreeable) model, and observe that many natural algorithms do not obtain constant-factor competitive ratios. In this section, we highlight several of these algorithms and give intuition for why they fail, even just for
unit-processing-time jobs. For each of the following examples, we assume that we have machine types $\{0,...,K\}$ (we have $K+1$ machine types in this section) with the capacity of each machine type $\ell$ being $10^{\ell}$ and the cost of each machine type
$\ell$ being $2^{\ell}$. We also assume all jobs have unit processing time.
Here, $K$ is an even integer that we will make large to obtain large
competitive ratios.

\paragraph*{Greedy}
Consider Algorithm \ref{algo:greedy}, which schedules all jobs that are currently present at each timeslot where some job has a deadline.
We showed in Section \ref{sec: agreeable} that this algorithm works well when the input jobs have agreeable deadlines.
In the general setting, the main problem with this algorithm is that it might have been better to schedule the majority of the jobs at or 
closer to their deadlines.
Suppose that at each timeslot in $\{1, 2, \ldots, 10^{K/2}\}$, a set of $10^{K/2}$ jobs arrive, of which one has an immediate deadline 
(i.e., at this timeslot) and all others have the same late deadline.
Greedy would schedule every job at its release time, in batches of $10^{K/2}$.
Meanwhile, \textsf{OPT} could use $10^{K/2}$ batches of type $0$ to schedule each job with an immediate deadline and one batch of cost $2^K$ to clear all jobs with the late deadline.
Then this algorithm incurs cost $10^{K/2} \cdot 2^{K/2}$, while \textsf{OPT} incurs cost $10^{K/2} + 2^{K}$.
The competitive ratio would be $\Omega(2^{K/2})$.

\paragraph*{Most cost-efficient}

Another algorithm is to use the machine with the best cost-per-job scheduled (after waiting for a deadline and using EDF).
The same setup used to show that Greedy can perform poorly also suffices to show that this cost-efficient algorithm performs poorly.
Using the same logic, we find that the cost-efficient algorithm incurs cost $10^{K/2} \cdot 2^{K/2}$, while \textsf{OPT} incurs cost $10^{K/2} + 2^{K}$.
The competitive ratio would again be $\Omega(2^{K/2})$.

\paragraph*{Lazy}

Since the Greedy algorithm was scheduling jobs too early, 
one might instead try a lazy approach in which the algorithm does the following: schedule the minimum number of jobs that have an immediate deadline, until there are enough jobs to fill the largest capacity machine, at which point use the largest capacity machine. One problem with this algorithm occurs when not enough jobs arrive to fill the largest machine. Suppose that $10^{K/2}$ jobs arrive at time $1$, each with a different deadline. Lazy would schedule each job separately on a machine of type $0$, while \textsf{OPT} would schedule all jobs at the beginning on a machine of type $K/2$. Then this algorithm incurs cost $10^{K/2}$, while \textsf{OPT} incurs cost $2^{K/2}$. The competitive ratio would be $\Omega(5^{K/2})$.

\paragraph*{Ramp-up}

We present one final simple algorithm we call the \emph{ramp-up algorithm}. 
It closely resembles deterministic
online algorithms designed for Multi Option Ski Rental
\cite{zhang2011ski,SLLA25}.
The ramp-up algorithm keeps track of an accumulated cost and always uses the largest machine type whose cost is at most the accumulated cost, or a machine of type $0$ if no
such machine exists. When the ramp-up algorithm clears all jobs, then it resets the accumulated cost to $0$. One problem is when the algorithm keeps ramping up and clearing too soon, it fails in the manner of the Greedy algorithm.

The following example shows that the ramp-up algorithm cannot obtain a constant factor competitive ratio.
Suppose that $2$ jobs arrive at time 0 and groups of $10^{\ell}$ jobs arrive at each succeeding timeslot 
$2 \ell + 1$, for $1 \leq \ell \leq K/2$.
At time $K+3$ one job arrives,
followed by one job at time $K+5$. 
In each group of jobs released at the same time,
one job has an immediate deadline and the rest have the same very late deadline.
Then this algorithm clears all waiting jobs in $3 + K/2$ 
machines for a total cost of at least $ 1 + 1 + \cdots + 2^{K/2} + 1 
\geq 2^{1 + K/2}$,
and resets the counter.
Suppose that this set of job arrivals happens $10^{(K/2)-1}$ times,
with staggered arrivals before the ``very late deadline". 
Then this algorithm would have a total cost of at least
$10^{(K/2) - 1}  \cdot 2^{1 + K/2}$,
while a solution of  total cost of $( 3 + K/2) \cdot 10^{(K/2) - 1} + 2^{K}$ exists:
use batches of type $0$  for the ``immediate" jobs from each group,
and one batch of type $K$ to schedule all the other jobs at the
common very late deadline.
The competitive ratio would be $\Omega(2^{K/2}/K)$.

Our Algorithm \ref{alg: pmain} does ``ramp up" to use bigger and bigger
machines, but manages to avoid the issue  of ramping up too fast; 
for example, for the kind of adversary provided in the previous paragraph, 
Algorithm \ref{alg: pmain} only uses machines of type $0$ for each group.

\section{Tight Example for the Lower Bound}\label{sec: app-tight-lb}

We show that the constant of $1/4$
given by Lemma \ref{lem: pvalid-assgn-lb} is tight, even for unit-processing-time
jobs. 
    Suppose we are given a set of unit jobs $J$ equipped with release
    times and deadlines in $[0,T]$. Let $\{(I,t_I, J_I)\}$ be a valid interval assignment $\mathcal{A}$ with $\cL$ the multiset of all intervals $I$ with $(I, t_I, J_I) \in \mathcal{A}$. 
For convenience, let
\[
    \sigma(\mathcal{A}) := \sum_{I \in \cL} 2^{t_I}.
\]

\begin{lemma}
    For any $\epsilon > 0$ there exists an instance and a valid
    interval assignment $\mathcal{A}$ such that
    $\text{cost}(\hOPT) <  \left(\frac14 + \epsilon \right) \sigma(\mathcal{A})$.
    \label{l_lb_tight}
\end{lemma}
\begin{proof}
Let $q$ be a large integer.
Suppose that machines of type $k \in \{0, 1, \ldots, q \}$
each have capacity $2^k$ and machines of type 
$k \in \{q+1, q+2, \ldots\}$ each have capacity $2^{q+k}$. Machines of type $k$  have cost $2^k$.

For this example, for every $I$ and every $j \in J_I$, $[r_j, d_j] = I$. 
We construct the following family of tuples. Figure \ref{f:tight} provides
an illustration for $q=3$.
For $r \in \{ 1, 2, 3, \ldots, 2^q \}$, and $s \in \{ 0, 1, \ldots, q+1 \}$,
we construct the tuple $Q_r^s = (I, t_I, J_I)$ where $I = [2r, 2r+1]$,
$t_I =s$, and $J_I$ is a set of ``new" jobs (a set disjoint from
the other sets constructed so far) of the size required by $t_I$.
That is, if $s=0$, then $|J_I| = 1$, and if $s \in \{ 1, 2, \ldots, q+1\}$,
then $|J_I| = 2^{s-1}$. We also have a tuple $Q' = (I, t_I, J_I)$
where $I = [0, 2^{q+1} + 2]$, $t_I = 2q+2$, and $J_I$ is
a set of new jobs of the required size, that is $|J_I| = 2^{3q+1}$.
These are all the jobs of this instance, and this is $\mathcal{A}$,
the family of tuples of this example.

\begin{figure}[th]
\centering
\begin{tabular}{c|c|c|c}
type $k$ & $c_k$ & $B_k$ & $|J_I|$ for $t_I = k$ \\
\hline
$0$ & $2^0 = 1$ & $2^0 = 1$ & 1 \\
\hline
$1$ & $2^1 = 2$ & $2^1 = 2$ & $B_{1-1} = 2^{1-1}  = 1$ \\
\hline
$2$ & $2^2 = 4$ & $2^2 = 4$ & $B_{2-1} = 2^{2-1} = 2$ \\
\hline
$3$ & $2^3 = 8$ & $2^3 = 8$ & $B_{3-1} = 2^{3-1} = 4$ \\
\hline
\hline
$4$ & $2^4 = 16$ & $2^7 = 128$ & $B_{4-1} = 2^{4-1} = 8$ \\
\hline
$5$ & $2^5 = 32$ & $2^8 = 256$ & $B_{5-1} = 2^{8-1} = 128$ \\
\hline
$6$ & $2^6 = 64$ & $2^9 = 512$ & $B_{6-1} = 2^{9-1} = 256$ \\
\hline
$7$ & $2^7 = 128$ & $2^{10} = 1024$ & $B_{7-1} = 2^{10-1} = 512$ \\
\hline
$8$ & $2^8 = 256$ & $2^{11} = 2048$ & $B_{8-1} = 2^{11-1} = 1024$ \\
\end{tabular}
 \vspace{0.2in}
\begin{center}\leavevmode%
\scalebox{0.5}{
  \includegraphics{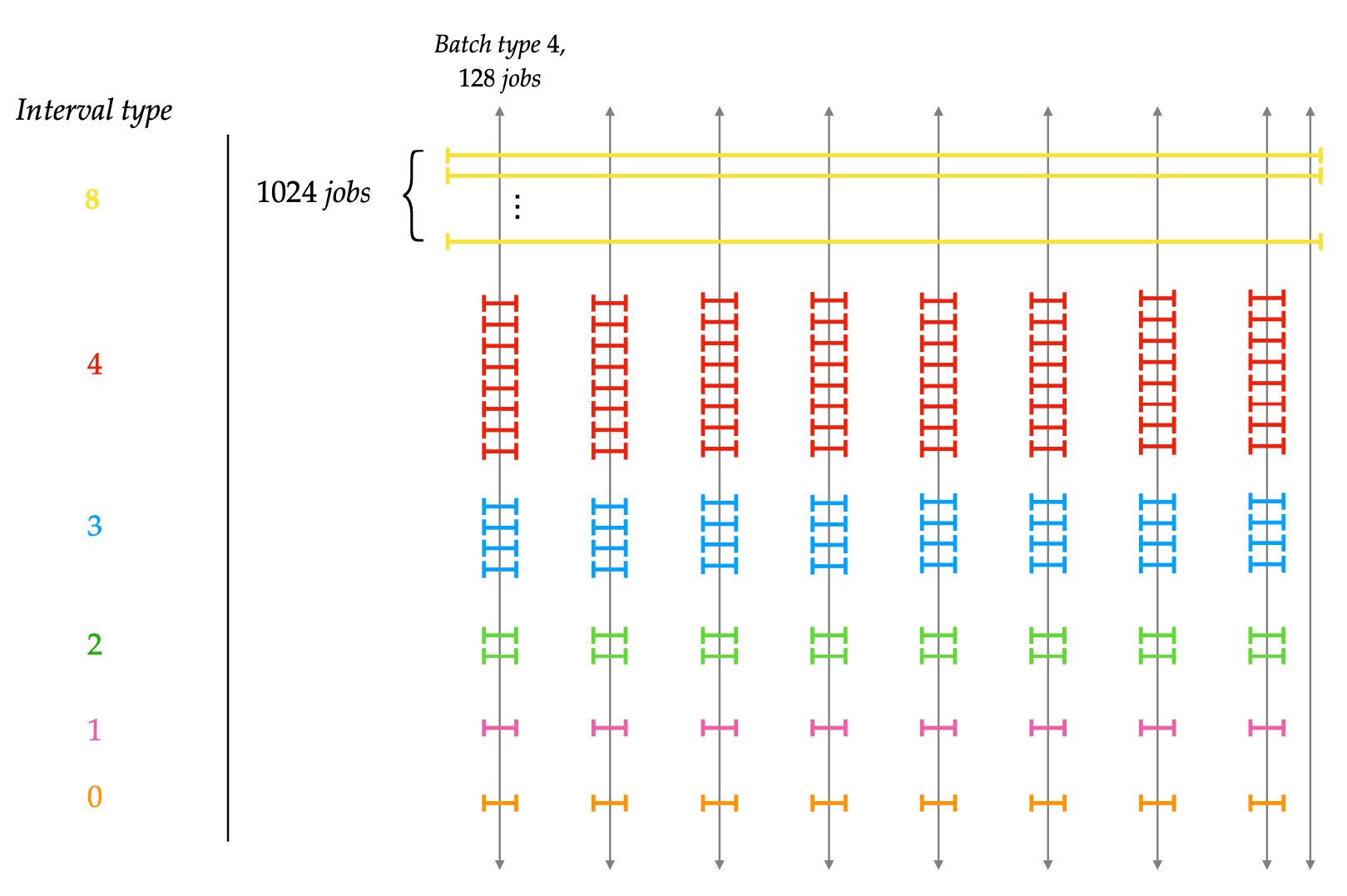}
}
\end{center}
\caption{The construction used in Lemma \ref{l_lb_tight} for $q=3$.
The family of tuples (comprising an interval assignment) $\mathcal{A}$ has one interval $I$ of type 8,
whose associated $J_I$ consists of 
the $1024$ yellow jobs, and for each $i = 0, \ldots, 4$, eight intervals $I$
of type $i$ with $J_I$ as depicted in the figure (8 red jobs for type 4, and so on).
Here a solution of cost $(2^3 + 1) 2^4 = 2^7 + 2^4$ exists: the $9$ vertical lines.
On the other hand, $\sigma(\mathcal{A}) = 2^3 ( 1 + 2 + 4 + 8 + 16) + 2^8 = 2^9 - 2^3$.
}
\label{f:tight}
\end{figure}

Then
\begin{equation*}
   \sigma(\mathcal{A}) =  2^q (2^0 + 2^1 + \cdots + 2^{q+1}) + 2^{2q+2}
   = 2^q ( 2^{q+2} - 1) + 2^{2q+2} = 2^{2q+3} - 2^q.
\end{equation*}

Now we construct a 
feasible schedule for this instance. 
For $r \in \{1, 2, 3, \ldots, 2^q\}$, we use a batch of type $q+1$ at time $2r+1$
to  schedule all the jobs assigned to $Q_r^s$, 
for all $s = 0, 1, \ldots, q+1$. There are exactly 
$ 1 + 1+  2 + 4 +  \cdots + 2^q = 2^{q+1}$ such jobs.
We fill this batch with $2^{2q+1} - 2^{q+1}$  jobs assigned to $Q'$.  
After doing this for all $r \in \{1 ,2, 3, \ldots, 2^q \}$, we have scheduled
all the jobs assigned to $Q_r^s$ for all $r$ and all $s$,
and we have scheduled $2^q ( 2^{2q+1}  - 2^{q+1} ) = 2^{3q+1} - 2^{2q+1}$
of the jobs assigned to $Q'$.
We use another batch of type $q+1$  at time $2^{q+1}+2$ to schedule the remaining $2^{2q+1}$
jobs that were assigned to $Q'$.
The cost of this solution is:
\[
\left( 2^q + 1 \right) 2^{q+1} = 2^{2q+1} + 2^{q+1}.
\]
By making $q$ large enough we finish the proof of the lemma.

\end{proof}

Therefore,
in order to get a competitive ratio better than 4 in the setting
where uniform-processing-time jobs have arbitrary intervals and costs 
that are powers of $2$, one cannot rely solely on the lower bound of
Lemma \ref{lem: pvalid-assgn-lb}.
Let us mention that for the Button problem or the Multi Option Ski Rental problem with costs powers of $2$,
deterministic 2-competitive online algorithms are possible.

\end{document}